\documentclass[aps,twocolumn,floats,pre]{revtex4}
\usepackage{graphics,graphicx,epsfig}
\usepackage{amssymb,color}
\usepackage{epsf,epstopdf,wrapfig}

\begin{document}

\title{Optimizing information flow in small genetic networks.}

\author{Ga\v{s}per Tka\v{c}ik,\footnote{gtkacik@sas.upenn.edu}$^a$ Aleksandra M. Walczak,\footnote{awalczak@princeton.edu}$^b$  and William Bialek\footnote{wbialek@princeton.edu}$^{b,c}$}

\affiliation{$^a$Department of Physics and Astronomy, University of Pennsylvania, Philadelphia, Pennsylvania 19104--6396\\
$^b$Joseph Henry Laboratories of Physics, Lewis--Sigler Institute for Integrative Genomics, and
Princeton Center for Theoretical Science,
Princeton University,
Princeton, New Jersey 08544\\
$^c$Center for Studies in Physics and Biology, The Rockefeller University, New York, New York 10065}

\date{\today}

\begin{abstract}
In order to survive, reproduce and (in multicellular organisms) differentiate, cells must control the concentrations of the myriad different proteins that are encoded in the genome.  The precision of this control is limited by the inevitable randomness of individual molecular events.  Here we explore how cells can maximize their control power in the presence of these physical limits; formally, we solve the theoretical problem of maximizing  the information transferred from inputs to outputs when the number of available molecules is held fixed.  We start with the simplest version of the problem, in which a single transcription factor protein controls the readout of one or more genes by binding to DNA.  We further simplify by assuming that this regulatory network operates in steady state, that the noise is small relative to the available dynamic range, and that the target genes do not interact.  Even in this simple limit, we find a surprisingly rich set of optimal solutions.  Importantly, for each locally optimal regulatory network, all parameters are determined once the physical constraints on the number of available molecules are specified.   Although we are solving an over--simplified version of the problem facing real cells, we see parallels between the structure of these optimal solutions and the behavior of actual genetic regulatory networks.  Subsequent papers will discuss more complete versions of the problem. 
\end{abstract}

\maketitle

\section{Introduction}

Much of the everyday business of organisms involves the transmission and processing of information.  On our human scale, the familiar examples involve the signals taken in through our sense organs \cite{spikes}.  On a cellular scale, information flows from receptors on the cell surface into the cell, modulating biochemical events and ultimately controlling gene expression \cite{ptashne+gann_02}.  In the course of development in multicellular organisms, individual cells acquire information about their location in the embryo  by responding to particular ``morphogen'' molecules whose concentration varies along the main axes of the embryo \cite{wolpert_69,lawrence_92}.  In all these  examples, information of interest to the organism ultimately is represented by events at the molecular level, whether the molecules are transcription factors regulating gene expression or ion channels controlling electrical signals in the brain.  This representation is limited by fundamental physical principles:  individual molecular events are stochastic, so that with any finite number of molecules there is a limit to the precision with which small signals can be discriminated reliably, and there is a limit to the overall dynamic range of the signals.   Our goal in this paper (and its sequel) is to explore these limits to information transmission in the context of small genetic control circuits.

The outputs of genetic control circuits are protein molecules that are synthesized by the cell from messenger RNA (mRNA), which in turn is transcribed from the DNA template.  The inputs often are protein molecules as well, ``transcription factors'' that bind to the DNA and regulate the synthesis of the mRNA.    In the last decade, a number of experiments have mapped the input/output relations of these regulatory elements, and characterized their noise, that is the fluctuations in the output protein concentration when the inputs are held fixed \cite{elowitz+al_02,ozbudak+al_02,blake+al_03,setty+al_03,raser+oshea_04,rosenfeld+al_05,pedraza+oudenaarden_05,golding+al_05,bar-even+al_06,newman+al_06,kuhlman+al_07,gregor+al_07a,gregor+al_07b}.  In parallel, a number of theoretical papers have tried to understand the origins of this noise, which ultimately reflects the random behavior of individual molecules along the path from input to output---the arrival of transcription factors at the their targets along the DNA, the initiation of transcription and the degradation of mRNA, the initiation of protein synthesis and the degradation of the output proteins 
\cite{arkin+al_98,paulsson_04,bialek+setayeshgar_05, bialek+setayeshgar_08, tkacik+bialek_07,tkacik+al_08a,tanase+al06,keplerelston01,swain+al_02,walczak+al05,buchler+al05,morelli+al08}.  While open questions remain, it seems fair to say that we have a physical picture of the noise in genetic control that we can use to ask questions about the overall function and design of these systems.

The ability of any system to transmit information is determined not just by input/output relations and noise levels, but also by the distribution of inputs; maximal information transmission requires a matching between the intrinsic properties of the system and the input statistics \cite{shannon_48,cover+thomas_91}.  In the context of sensory information processing, these matching conditions have been explored almost since the inception of information theory \cite{attneave_54,barlow_59,laughlin_81,atick+redlich_90}.  In particular, because the distribution of sensory inputs varies with time, optimal information transmission requires that the input/output relation track or adapt to these variations, and this theoretical prediction has led to a much richer view of adaptation in the neural code \cite{smirnakis+al_97,brenner+al_00,fairhall+al_01,maravall+al_07,wark+al_07}.  There are analogous matching conditions for genetic regulatory elements, and these conditions provide parameter free predictions about the behavior of the system, based on the idea that cells are trying to transmit the maximum amount of information \cite{tkacik+al_08b}.  Comparison with recent experiments has been encouraging \cite{tkacik+al_08c}.

In this paper we go beyond the matching conditions to ask how cells can adjust the input/output relations of genetic regulatory elements so as to maximize the information that is transmitted through these systems.  Absent any constraints, the answer will always be to make more molecules, since this reduces the effective noise level, so we consider the problem of maximizing information transmission with a fixed mean or maximum number of molecules at both the input and the output.  In this sense we are asking how cells can extract the maximum control power, measured in bits, from a given number of molecules, thus optimizing functionality under clear physical constraints.  In general this problem is very difficult, so we start here with the simplest case of a single input transcription factor that controls (potentially) many genes, but there is no interaction among these outputs.  Further, we focus on a limit (small noise) where some analytic progress is possible.  We will see that, even in this case, the optimal solutions have an interesting structure, {\color{black} which emerges as a result of the  interplay  between   noise sources at the input and the output of the regulatory elements.} For other approaches to the optimization of information transmission in biochemical and genetic networks, see Refs \cite{ziv+al_07,emberly_08,tostvein+tenwolde_09}.

Optimization of information transmission is a concise, abstract principle, grounded in the physics of the molecular interactions that underlie biological function.  It would be attractive if we could derive the behavior of biological systems from such a principle, rather than taking the myriad parameters of these systems simply as quantities that must be fit to data.   It is not at all clear, however, that such a general principle should apply to real biological systems.  Indeed, it is possible that solutions to our optimization problem are far from plausible in comparison with what we find in real cells.
Thus, our most important result is that the parameters which we derive are reasonable in relation to experiment.  While a realistic comparison requires us to solve the optimization problem in a fully interacting system,   even in the simpler problem discussed here we can see the outlines of a theory for real genetic networks. {\color{black} Subsequent papers will address the full, interacting version of the problem. }

\section{Formulating the problem}

A gene regulatory element translates the concentration of input molecules $\mathcal{I}$ into output molecules $\mathcal{O}$.   We would like to measure, quantitatively, how effectively changes in the input serve to control the output.  If we make many observations on the state of the cell, we will see that inputs and outputs are drawn from a joint distribution $p(\mathcal{I},\mathcal{O})$, and our measure of control power should be a functional of this distribution.   
In his classic work, Shannon showed that there is only one such measure of control power which obeys certain plausible constraints, and this is the mutual information between $\cal I$ and $\cal O$   \cite{shannon_48,remark}.

To be concrete, we consider a 
set of genes, ${\rm i} = 1, 2, \cdots , M$, that all are controlled by a single transcription factor.  Let the concentration of the transcription factor be $c$ and let the levels of protein expressed from each gene be $g_{\rm i}$; below we discuss the units and normalization of these quantities.    Thus, the input ${\cal I} \equiv c$ and the output ${\cal O} \equiv \{g_{\rm i}\}$.  
{\color{black}In principle these quantities all depend on time. We choose to focus here on the steady state problem, where we assume that the output expression levels reach their equilibrium values before the input transcription factor concentrations change.  

We view the steady state approximation not necessarily as an accurate model of the dynamics in real cells, but as a useful starting point, and already the steady state problem has a rich structure.  In particular, as we will see, in this limit we have analytic control over the role of nonlinearities in the input/output relation describing the function of the different regulatory elements in our network.  In contrast, most approaches to information transmission by dynamic signals are limited to the regime of linear response; see, for example, Ref \cite{tostvein+tenwolde_09}.    Although we are focused here on information transmission in genetic circuits, it is interesting that the same dichotomy---nonlinear analyses of static networks and dynamic analyses of linear networks---also exists in the literature on information transmission in neural networks \cite{laughlin_81,atick+redlich_90}.}


To specify the joint distribution of inputs and outputs, it is convenient to think that 
the transcription factor concentration is being chosen out of a probability distribution $P_{TF}(c)$, and then the target genes respond with expression levels chosen out of the conditional distribution $P(\{g_{\rm i}\}| c)$.  In general, the mutual information between the set of expression levels $\{ g_{\rm i}\}$ and the input $c$ is given by \cite{shannon_48,cover+thomas_91}
\begin{widetext}
\begin{equation}
I(\{g_{\rm i}\}; c) = \int dc \,  \int d^M g \, P(c, \{g_{\rm i}\})
\log_2 \left[
{{P(c, \{g_{\rm i}\})}\over {P_{TF}(c) P(\{g_{\rm i}\})}}
\right] \, {\rm bits} ,
\end{equation}
\end{widetext}
where the overall distribution of expression levels is given by
\begin{equation}
P(\{ g_{\rm i}\}) = \int dc\, P_{TF}(c)P(\{g_{\rm i}\}| c) .
\end{equation}
Shannon's uniqueness theorem of course leaves open a choice of units, and here we make the conventional choice of bits, hence the logarithm is base two.


We will approach the problem of optimizing information transmission in two steps. First, we will adjust the distribution $P_{TF}(c)$ to take best advantage of the input/output relations, and then we will adjust the input/output relations themselves.  Even the first step is difficult in general, so we start by focusing on the limit in which noise is small.

\subsection{Information in the small noise limit}
As noted in the Introduction, we will confine our attention in this paper to the case where each gene responds independently to its inputs, and there are no interactions among the output genes; we point toward generalizations in the Discussion below, and return to the more general problem in subsequent papers.  The absence of interactions means that the conditional distribution of expression levels must factorize,
$P(\{g_{\rm i}\}|c) = \prod_{{\rm i}=1}^M P_{\rm i} (g_{\rm i} | c) $.
Further, we assume that the noise in expression levels is Gaussian.  Then we have \cite{note_Gauss}
\begin{eqnarray}
P(\{g_{\rm i}\}| c)  &=&  \exp{\bigg[} - {M\over 2}\ln (2\pi )
- {1\over 2} \sum_{{\rm i}=1}^M\ln ( {\sigma_{\rm i}^2(c)}) \nonumber\\
&&\,\,\,\,\,\,\,\,\,\,\,\,\,\,\, - {1\over 2} \sum_{{\rm i}=1}^M  \frac{1}{\sigma_{\rm i}^2(c)}
\left( g_{\rm i} - \bar g_{\rm i}(c)\right)^2 
{\bigg]} .
\end{eqnarray}
The input/output relation of each gene is defined by the mean $\bar g_{\rm i}(c)$, while $\sigma_i^2$ measures the variance of the fluctuations or noise in the expression levels at fixed input,
\begin{equation}
\sigma_{\rm i}^2(c) = \langle (g_{\rm i} - \bar g_{\rm i}(c))^2\rangle .
\end{equation}

In the limit that the noise levels $\sigma_{\rm i}$ are small, we can develop a systematic expansion of the information $I(\{g_{\rm i}\}; c)$, generalizing the approach of Refs \cite{tkacik+al_08b,tkacik+al_08c}.  The key idea is that, in the small noise limit, observation of the output expression levels $\{g_{\rm i}\}$ should be sufficient to determine the input concentration $c$ with relatively high accuracy; further, we expect that errors in this estimation process would be well approximated as Gaussian.  Formally, this means that we should have
\begin{equation}
P(c|\{g_{\rm i}\}) \approx {1\over{\sqrt{2\pi \sigma_c^2 (\{g_{\rm i}\})}}}
\exp\left[ -{{(c - c^*(\{g_{\rm i}\}))^2}\over {2\sigma_c^2 (\{g_{\rm i}\})}}\right] ,
\end{equation}
where $ c^*(\{g_{\rm i}\})$ is the most likely value of $c$ given the outputs, and $\sigma_c^2 (\{g_{\rm i}\})$ is the variance of the true value around this estimate.  
We can use this expression to calculate the information by writing $I(\{g_{\rm i}\};c)$ as the difference between two entropies:
\begin{widetext}
\begin{eqnarray}
I(\{g_{\rm i}\};c) 
&=& - \int dc\,P_{TF}(c)\log_2 P_{TF}(c) -  \int d^M g\, P(\{g_{\rm i}\}) \left[ - \int dc \, P(c|\{g_{\rm i}\})\log_2 P(c|\{g_{\rm i}\})\right] \\
&=& - \int dc\,P_{TF}(c)\log_2 P_{TF}(c) - {1\over 2} \int d^M g\, P(\{g_{\rm i}\})\log_2\left[ 2\pi e \sigma_c^2 (\{g_{\rm i}\})\right] .
\end{eqnarray}
\end{widetext}
Intuitively, the first term is the entropy of inputs, which sets an absolute maximum on the amount of information that can be transmitted \cite{continuous}; the second term is (minus) the entropy of the input given the output, or the ``equivocation'' \cite{shannon_48} that results from noise in the mapping from inputs to outputs.
To complete the calculation we need an expression for this effective noise level $\sigma_c$.  

Using Bayes' rule, we have
\begin{eqnarray}
P(c|\{g_{\rm i}\})  &=& {{P(\{g_{\rm i}\}|c ) P_{TF}(c)}\over{P(\{g_{\rm i}\})}}\\
&=& {1\over{{\cal Z}(\{g_{\rm i}\})}} \exp\left[ - F(c, \{g_{\rm i}\})\right] ,
\end{eqnarray}
where 
\begin{eqnarray}
F(c, \{g_{\rm i}\}) &=& -\ln P_{TF}(c) + {1\over 2} \sum_{{\rm i}=1}^M\ln ( {\sigma_{\rm i}^2(c)}) \nonumber \\
&&+ {1\over 2} \sum_{{\rm i}=1}^M  \frac{1}{\sigma_{\rm i}^2 (c)}
\left( g_{\rm i} - \bar g_{\rm i}(c)\right)^2 .
\end{eqnarray}
Now it is clear that $c^*(\{g_{\rm i}\})$ and $\sigma_c(\{g_{\rm i}\})$ are defined by
\begin{eqnarray}
0 &=& {{\partial F(c, \{g_{\rm i}\})}\over{\partial c}} {\Bigg |}_{c = c^*(\{g_{\rm i}\})} ,\\
{1\over{\sigma_c^2(\{g_{\rm i}\})}} &=& {{\partial^2 F(c, \{g_{\rm i}\})}\over{\partial c^2}} {\Bigg |}_{c = c^*(\{g_{\rm i}\})} .
\end{eqnarray}
The leading term at small $\sigma_{\rm i}$ is then given by
\begin{equation}
{1\over {\sigma_c^2 (\{g_{\rm i}\})}} = \sum_{{\rm i}=1}^M 
\frac{1}{\sigma_{\rm i}^2}
\left(
{{d\bar g_{\rm i}(c)}\over{dc}} \right)^2 
{\Bigg |}_{c = c^*(\{g_{\rm i}\})} .
\end{equation}

Finally, we note that, in the small noise limit, averages over all the expression levels can be approximated by an integral along the trajectory of mean expression levels, with an appropriate Jacobian.  More precisely,
\begin{equation}
\int d^M g \, P(\{ g_{\rm i}\}) \,\left[ \cdots \right]
 \approx 
\int dc \, P_{TF}(c) \prod_{{\rm i}=1}^M \delta (g_{\rm i} - \bar g_{\rm i}(c))) \,\left[ \cdots \right] .
\end{equation}
Putting all these terms together, we have
\begin{widetext}
\begin{equation}
I(\{g_{\rm i}\}; c) = -\int dc\, P_{TF}(c) \log_2 P_{TF}(c) + {1\over 2}
\int dc\, P_{TF}(c) \log_2\left[{1\over{2\pi e}} \sum_{{\rm i}=1}^M 
\frac{1}{\sigma_{\rm i}^2(c)}
\left(
{{d\bar g_{\rm i}(c)}\over{dc}} \right)^2 
\right] . \label{eqinfo}
\end{equation}
\end{widetext}

The small noise approximation is not just a theorist's convenience.  A variety of experiments show that fluctuations in gene expression level can be $10-25\%$ of the mean \cite{elowitz+al_02,raser+oshea_04,rosenfeld+al_05,newman+al_06,bar-even+al_06,gregor+al_07b}. As noted above, maximizing information transmission requires matching the distribution of input signals to the structure of the input/output relations and noise, and in applying these conditions to a real regulatory element in the fruit fly embryo it was shown that the (analytically accessible) small noise approximation gives results which are in semi--quantitative agreement with the 
(numerical) exact solutions \cite{tkacik+al_08c}.  Thus, although it would be interesting to explore the quantitative deviations from the small noise limit, we believe that this approximation is a good guide to the structure of the full problem.

{\color{black}  To proceed, Eq (\ref{eqinfo}) for the information in the small noise limit instructs us to compute the mean response, $\bar{g}_i(c)$ and the noise, $\sigma_i(c)$, for every regulated gene. Since the properties of noise in gene expression determine to a large extent  the structure of optimal solutions, we   present  in Sec \ref{ss:io}. a detailed description of these noise sources.  In Sec \ref{ss:cst} we then introduce the `cost of coding,' measured by the number of signaling molecules that the cell has to pay to transmit the information reliably. Finally, we look for optimal solutions in Sec \ref{ss:res}.} 

\subsection{Input/output relations and noise}
\label{ss:io}

Transcription factors act by binding to DNA near the point at which the ``reading'' of a gene begins, and either enhancing or inhibiting the process of transcription into mRNA. 
\begin{figure}[b]
\includegraphics[width = \linewidth]{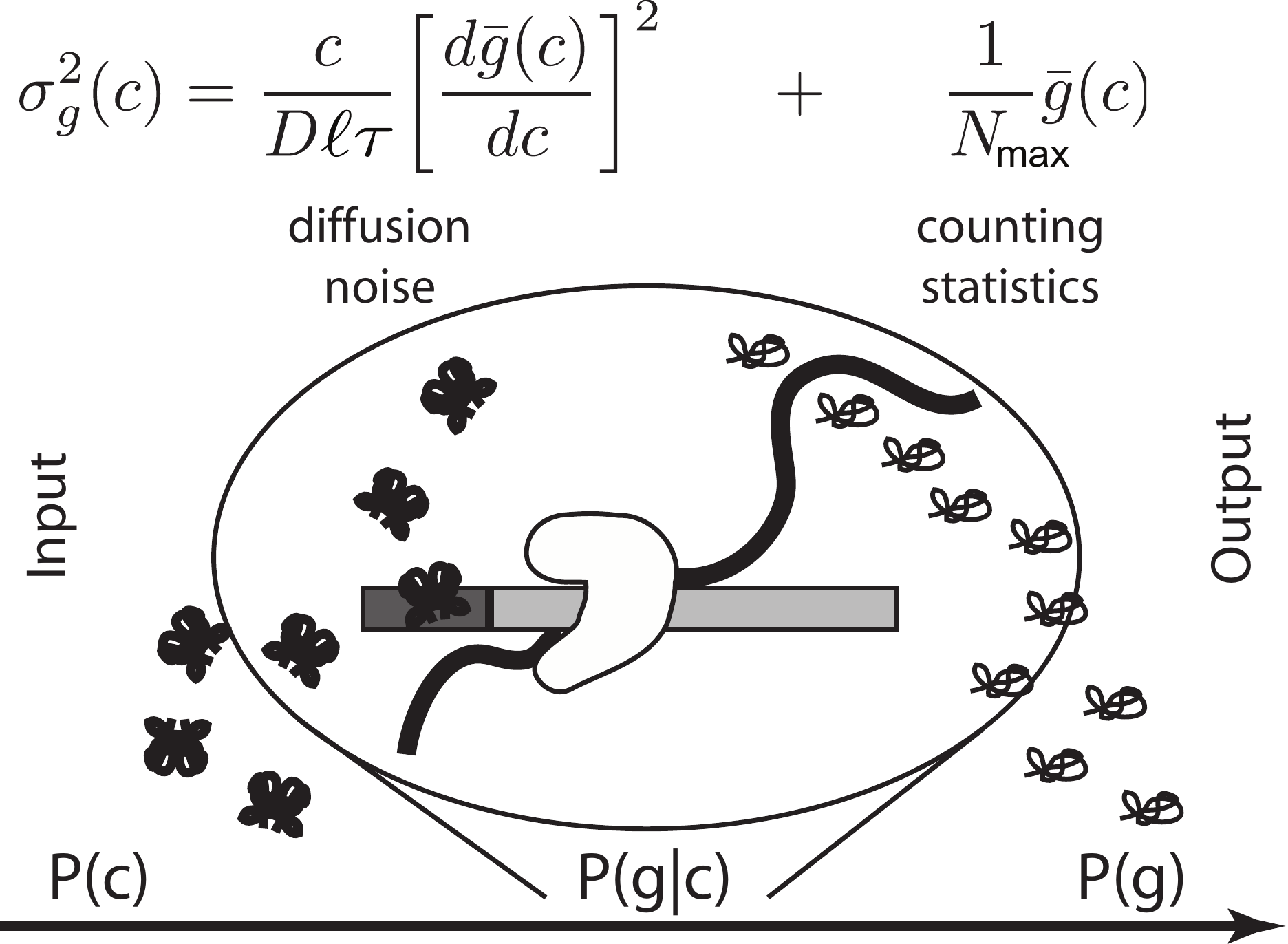}
\caption{Input proteins at concentration $c$ act as transcription factors for the expression of output proteins, $g$. The diffusive noise in transcription factor concentration  and the shot noise at the output both contribute to stochastic gene expression. The regulation process is described using a conditional probability distribution of the output knowing the input, $P(g|c)$, which can be modeled as a Gaussian process with a variance $\sigma^2_g(c)$. In this paper we consider the case of multiple output genes $\left\{g_{\rm i}\right\}$, $i=1,\cdots,M$, each of which is independently regulated by the process illustrated here with the corresponding noise $\sigma_{\rm i}^2$. }
\label{graph}
\end{figure}
In bacteria, a simple geometrical view of this process seems close to correct, and one can try to make a detailed model of the energies for binding of the transcription factor(s) and the interaction of these bound factors with the transcriptional apparatus, RNA polymerase in particular \cite{bintu+al_05a,bintu+al_05b}.  For eukaryotes the physical picture is less clear, so we proceed phenomenologically.  If binding of the transcription factor activates the expression of gene $\rm i$, we write
\begin{equation}
\bar g_{\rm i} (c) = \frac{c^{n_{\rm i}}}{c^{n_{\rm i}}+{K_{\rm i}}^{n_{\rm i}}} ,
\end{equation}
and similarly if the transcription factor represses expression we write
\begin{equation}
\bar g_{\rm i} (c) = \frac{K_{\rm i}^{n_{\rm i}}}{c^{n_{\rm i}}+{K_{\rm i}}^{n_{\rm i}}} .
\end{equation}
These are smooth, monotonic functions that interpolate between roughly linear response ($n=1$ and large $K$) and steep, switch--like behavior ($n\rightarrow\infty$) at some threshold concentration ($c=K$).  Such `Hill functions' often are used to describe the cooperative binding of $n$ molecules to their target sites \cite{hill}, with $F = -k_B T \ln K$ the free energy of binding per molecule, and this is a useful intuition even if it is not correct in detail.

To complete our formulation of the problem we need to understand the noise or fluctuations in expression level at fixed inputs, as summarized by variances ${\sigma}_{\rm i}^2$. 
There are several contributions to the variance, which we can divide into two broad categories, as in Fig \ref{graph}.

The transcription of mRNA and its translation into protein can be thought of as the ``output'' side of the regulatory apparatus.  Ultimately these processes are composed of individual molecular events, and so there should be shot noise from the inherent randomness of these events.  This suggests that there will be an output noise variance proportional to the mean, $\sigma_{\rm i,\, out}^2 \propto \bar g_{\rm i}$.

The arrival of transcription factor molecules can be thought of as the ``input'' side of the apparatus, and again there should be noise associated with the randomness in this arrival.  This noise is equivalent to a fluctuation in the input concentration itself; the variance in concentration should again be proportional to the mean, and the impact of this noise needs to be propagated through the input/output relation, so that
$\sigma_{\rm i,\, in}^2 \propto c (d\bar g_{\rm i}/dc)^2$.

Putting together the input and output noise, we have
\begin{equation}
\sigma_{\rm i}^2 (c) = a \bar g_{\rm i} (c) + b c \left( {{d\bar g_{\rm i}(c)}\over{dc}}\right)^2 ,
\label{noise1}
\end{equation}
where $a$ and $b$ are constants.  Comparing this intuitive estimate to more detailed calculations \cite{bialek+setayeshgar_05, tkacik+al_08a}
allows us to interpret these constants.  If $\bar g_{\rm i}$ is normalized so that its maximum value is one, then $a = 1/N_{\rm max}$, where $N_{\rm max}$ is the maximum number of {\rm independent} molecules that are made from gene $\rm i$.  If, for example, each mRNA molecule generates many proteins during its lifetime, then (if the synthesis of mRNA is limited by a single kinetic step) $N_{\rm max}$ is the maximum number of mRNAs, as discussed in Refs \cite{swain+al_02,paulsson_04,tkacik+al_08a}.

The shot noise in the arrival of transcription factors at their targets ultimately arises from diffusion of these molecules.  Analysis of the coupling between diffusion and the events that occur at the binding site \cite{bialek+setayeshgar_05,bialek+setayeshgar_08,tkacik+bialek_07} shows that the total input noise has both a term $\propto c (d\bar g_{\rm i}/dc)^2$ and additional terms that can be made small by adjusting the parameters describing kinetics of steps that occur after the molecules arrive at their target; here we assume that Nature chooses parameters which make these non--fundamental noise sources negligible \cite{note2}.  In the remaining term, we have $b \sim 1/(D\ell \tau)$, where $D$ is the diffusion constant of the transcription factor, $\ell$ is the size of its target on the DNA, and $\tau$ is the time over which signals are integrated in establishing the steady state.

\begin{table}[b]
\begin{center}
\begin{tabular}{|c|c|c|c|}
\hline
concentration & scale & system & Ref\\\hline
$55\pm 10\, {\rm nM}$ & midpoint & $\lambda$ repressor in {\em E coli} & \cite{rosenfeld+al_05}\\\hline
$55\pm 3 \,{\rm nM}$ & maximum & Bcd in {\em Drosophila} embryo & \cite{gregor+al_07b} \\\hline
$5.3\pm 0.7\,{\rm nM}$ & midpoint&GAGA   & \cite{pedone+al96} \\\hline
$\sim 5\,{\rm nM}$ &  midpoint & crp to {\em lac} site & \cite{bintu+al_05b} \\\hline
$\sim 0.2\,{\rm nM}$ &midpoint & lac to {OR1} & \cite{bintu+al_05b, oehler+al_94} \\\hline
$\sim 3 \,{\rm nM}$ & midpoint &  {lac} to {OR2} & \cite{bintu+al_05b, oehler+al_94} \\\hline
$\sim 110 \,{\rm nM}$ & midpoint &   lac to {OR3} & \cite{bintu+al_05b, oehler+al_94} \\\hline
$22\pm 3\,{\rm nM}$ & midpoint &  { lac} to {OR1} in vitro & \cite{wang+al_08} \\\hline
\end{tabular}
\end{center}
\caption{Concentration scales for transcription factors.  We collect absolute concentration measurements on transcription factors from several different systems, sometimes indicating the maximum observed concentration and in other cases the concentration that achieves half--maximal activation or repression (midpoint). Bcd is the bicoid protein, a transcription factor involved in early embryonic pattern formation; GAGA is a transcription factor in {\em Drosophila}, crp is a transcription factor that acts on a wide range of metabolic genes in bacteria; {\em lac} is the well studied operon that encodes proteins needed for lactose metabolism in {\em E coli}; lac is the transcription factor that represses expression of the {\em lac} operon; OR1--3 are binding sites  for the lac repressor.
\label{Cs}}
\end{table}%

With the (semi--)microscopic interpretation of the parameters, we can write
\begin{equation}
\sigma_{\rm i}^2 (c) = {1\over{N_{\rm max}}}
\left[ 
\bar g_{\rm i} (c) +  c c_0 \left( {{d\bar g_{\rm i}(c)}\over{dc}}\right)^2 \right],
\label{noiseform}
\end{equation}
where there is a natural scale of concentration,
\begin{equation}
c_0 = {{N_{\rm max}}\over{D \ell \tau}} .
\label{c0}
\end{equation}
To get a rough feeling for this scale,
we note that diffusion constants for proteins in the cytoplasm are $\sim \mu {\rm m}^2 /{\rm s}$ \cite{elowitz+al_99,goldingcox06,gregor+al_07a,elf+al_07}, target sizes are measured in $\rm nm$, and integration times are minutes or hundreds of seconds (although there are few direct measurements).  The maximum number of independent molecules depends on the character of the target genes.  In many cases of interest, these are also transcription factors, in which case a number of experiments suggest that $N_{\rm max} \sim 10 -100$ 
\cite{golding+al_05,paulsson_04,tkacik+al_08a}.  
Putting these numbers together, we have $c_0 \sim 10-100/(\mu{\rm m})^3$ or $\sim 15 - 150\,{\rm nM}$, although this (obviously) is just an order of magnitude estimate.

{\color{black} To summarize, two rather general forms of noise limit the information transmission in genetic regulatory networks. Both combine additively and ultimately trace their origin to a finite (and possibly small) number of signaling molecules. The input noise is caused by a small concentration of transcription factor molecules, and its effect on the regulated gene is additionally modulated by the input--output relation. The output noise is caused by the small number of gene products, and this noise is simply proportional to the mean. It is reasonable to believe that the strengths of these two noise sources, in appropriate units, will be of comparable magnitude. Since the organism has to pay a metabolic price to reduce either noise source, it would be wasting resources if it were to lower the strength of one source alone far below the limiting effect of the other.}

\subsection{Constraining means or maxima}
\label{ss:cst}

To proceed, we need to decide how the problem of maximizing information transmission will be constrained.  One possibility is that we fix the maximum number of molecules at the input and the output.  The constraint on the output can be implemented by measuring the expression levels in units such that the largest values of the mean expression levels $\bar g_{\rm i}$ are all equal to one \cite{note1}.  On the input side, we restrict the range of $c$ to be $c\in [0 ,c_{\rm max}]$.
With this normalization and limits on the $c$ integrals, we can maximize $I(\{g_{\rm i}\}; c)$ directly by varying the distribution of inputs, adding only a Lagrange multiplier to fix the normalization of $P_{TF}(c)$,
\begin{equation}
{\delta\over{\delta P_{TF}(c)}}\left[ I(\{g_{\rm i}\}; c)
- \lambda \int dc\, P_{TF}(c)\right] = 0 .
\label{var1}
\end{equation}

As discussed in Ref \cite{tkacik+al_08c}, 
the solution to the variational problem defined in Eq (\ref{var1}) is
\begin{eqnarray}
P_{TF}^*(c) &=& \frac{1}{Z_1\sqrt{2\pi e}}\frac{1}{\sigma_c}\\
&=& {1\over {Z_1}}
\left[{1\over{2\pi e}} \sum_{{\rm i}=1}^M 
\frac{1}{\sigma_{\rm i}^2(c)}
\left(
{{d\bar g_{\rm i}(c)}\over{dc}} \right)^2 
\right]^{1/2}  , 
\end{eqnarray}
where the normalization constant $Z_1$ is given by
\begin{equation}
Z_1 = \int_0^{c_{\rm max}} dc\, \left[{1\over{2\pi e}} \sum_{{\rm i}=1}^M
\frac{1}{\sigma_{\rm i}^2(c)}
\left(
{{d\bar g_{\rm i}(c)}\over{dc}} \right)^2 
\right]^{1/2}   .
\label{Z1_multiple}
\end{equation}
The information transmission with this optimal choice of $P_{TF}(c)$ takes a simple form, 
\begin{equation}
I_1^* = \log_2 Z_1 .
\label{infofromZ1}
\end{equation}

The expression for $Z_1$, and hence the optimal information transmission, has a simple geometric interpretation.  As the concentration of the input transcription factor varies, the output moves, on average, along  a trajectory in the $M$--dimensional space of expression levels; this trajectory is defined by $\{\bar g_{\rm i}(c)\}$.  Nearby points along this trajectory can't really be distinguished, because of noise; the information transmission should be related to the number of distinguishable points.  If the noise level were the same everywhere,  this count of distinguishable states would be just the length of the trajectory in units where the standard deviation of the output fluctuations, projected along the trajectory, is one.  Since the noise isn't uniform, we should introduce the local noise level into our metric for measuring distances in the space of expression levels, and this is exactly what we see in Eq (\ref{Z1_multiple}). Thus, we can think of the optimal information transmission as being determined by the length of the path in expression space that the network traces as the input concentration varies, where length is measured with a metric determined by the noise level.  

This information capacity still depends upon the input/output relations and the noise levels, so we have a second layer of optimization that we can perform.  Before doing this, however, we consider another formulation of the constraints.

As an alternative to fixing the maximum concentration of input transcription factor molecules, we consider fixing the mean concentration.  To do this, we introduce, as usual, a second Lagrange multiplier $\alpha$, so that our optimization problem becomes
\begin{widetext}
\begin{equation}
{\delta\over{\delta P_{TF}(c)}}\left[ I(\{g_{\rm i}\}; c)
- \lambda \int dc\, P_{TF}(c)
-\alpha \int dc\, P_{TF}(c) c \right] = 0 .
\label{optmean1}
\end{equation}
\end{widetext}
Notice that we can also think of this as maximizing information transmission in the presence of some fixed cost per input molecule.

Solving Eq (\ref{optmean1}) for the distribution of inputs, $P_{TF}(c)$, we find
\begin{equation}
P_{TF}^*(c) = {1\over {Z_2}}
\left[{1\over{2\pi e}} \sum_{{\rm i,j}=1}^M 
\frac{1}{\sigma_{\rm i}^2(c)}
\left(
{{d\bar g_{\rm i}(c)}\over{dc}} \right)^2 
\right]^{1/2} e^{ -\alpha c } ,
\end{equation}
where
\begin{equation}
Z_2  = \int_0^\infty dc\, 
\left[{1\over{2\pi e}} \sum_{{\rm i,j}=1}^M
\frac{1}{\sigma_{\rm i}^2(c)}
\left(
{{d\bar g_{\rm i}(c)}\over{dc}} \right)^2 
\right]^{1/2} e^{ -\alpha c}  .
\label{Z2def}\end{equation}
As usual in such problems we need to adjust the Lagrange multipliers to match the constraints, which is equivalent to solving 
\begin{equation}
-{{\partial \ln Z_2}\over{\partial\alpha}} = \langle c \rangle.
\end{equation}
The optimal information transmission in this case is 
\begin{equation}
I_2^* = \log_2 Z_2 + \alpha \langle c \rangle .
\label{I2def}
\end{equation}

One might think that, for symmetry's sake, we should consider a formulation in which the mean number of output molecules also is constrained.  There is some subtlety to this, since if we know the input/output functions, $\{\bar g_{\rm i}(c)\}$, and the distribution of inputs, $P_{TF}(c)$, then the mean output levels are  determined.  Thus it is not obvious that we have the freedom to adjust the mean output levels.  We return to this point  in Section \ref{sec_conmean}.

\section{One input, one output}
\label{ss:res}

To get a feeling for the structure of our optimization problem, we consider the case where the transcription factor regulates the expression level of just one gene.  If we constrain the maximum concentrations at the input and output, then the information capacity is set by $I = \log_2 Z_1$ [Eq (\ref{infofromZ1})]; substituting our explicit expression for the noise [Eq (\ref{noiseform})] we have
\begin{equation}
Z_1 = \int_0^{c_{\rm max}} dc\, \left[ {{N_{\rm max}}\over {2\pi e}}
{{(d\bar g(c)/dc)^2}
\over{\bar g(c) + c_0 c (d\bar g(c)/dc)^2}
}
\right]^{1/2} .
\end{equation}

\begin{figure}
\includegraphics[width = \linewidth]{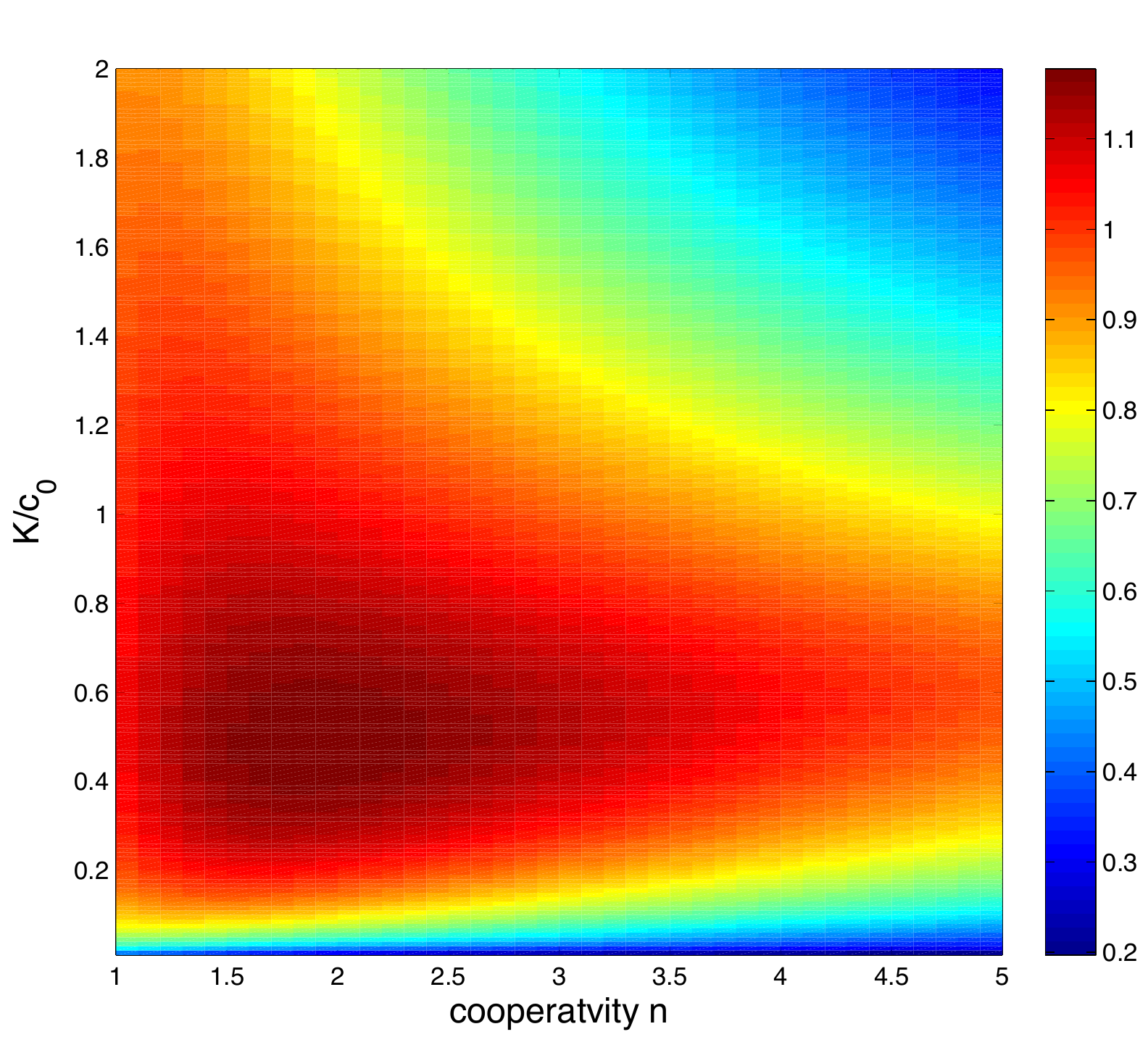}
\caption{(Color online) Information capacity for one (activator) input and one output. The information is $I=\log_2 \tilde Z_1 + A$, with $A$ independent of the parameters; the map shows $\tilde Z_1$ as computed from Eq (\ref{normZ1}), here with $C\equiv c_{\rm max}/c_0 = 1$.  We see that there is a broad optimum with cooperativity $n_{\rm opt} = 1.86$ and $K_{\rm opt} = 0.48 c_0 = 0.48c_{\rm max}$.  \label{Zmap1}}
\end{figure}

The first point to note is that if the natural scale of concentration, $c_0$, is either very large or very small, then the optimization problem loses all of its structure.  Specifically, in these two limits we have
\begin{eqnarray}
Z_1(c_0\rightarrow \infty ) &=& \left[ {{D\ell\tau}\over {2\pi e}}\right]^{1/2}
\int_0^{c_{\rm max}} {{dc}\over{\sqrt{c}}} ,\\
&=&  \left[ {{2D\ell\tau c_{\rm max}}\over {\pi e}}\right]^{1/2} ,
\end{eqnarray}
and 
\begin{eqnarray}
Z_1(c_0\rightarrow 0 ) &=& \left[ {{N_{\rm max}}\over {2\pi e}}\right]^{1/2}
\int_0^{c_{\rm max}} {{dc}\over{\sqrt{\bar g (c)}}} {\Bigg |}
{{d\bar g(c)}\over {dc}}{\Bigg |},\\
&=&  \left[ {{2N_{\rm max}}\over {\pi e}}\right]^{1/2}
{\bigg |} \sqrt{\bar g (c_{\rm max})} - \sqrt{\bar g (0)}{\bigg |} .\nonumber\\
&&
\label{smallc0}
\end{eqnarray}
In both cases, the magnitude of the information capacity becomes independent of the shape of the input/output relation $\bar g(c)$.  Thus, the possibility that real input/output relations are determined by the optimization of information transmission depends on the scale $c_0$ being comparable to the range of concentrations actually used in real cells.  Although we have only a rough estimate of $c_0\sim 15-150\,\mathrm{nM}$, Table \ref{Cs} shows that this is the case.

\subsection{Numerical results with $c_{\rm max}$}

\begin{figure}[b]
\includegraphics[width = \linewidth]{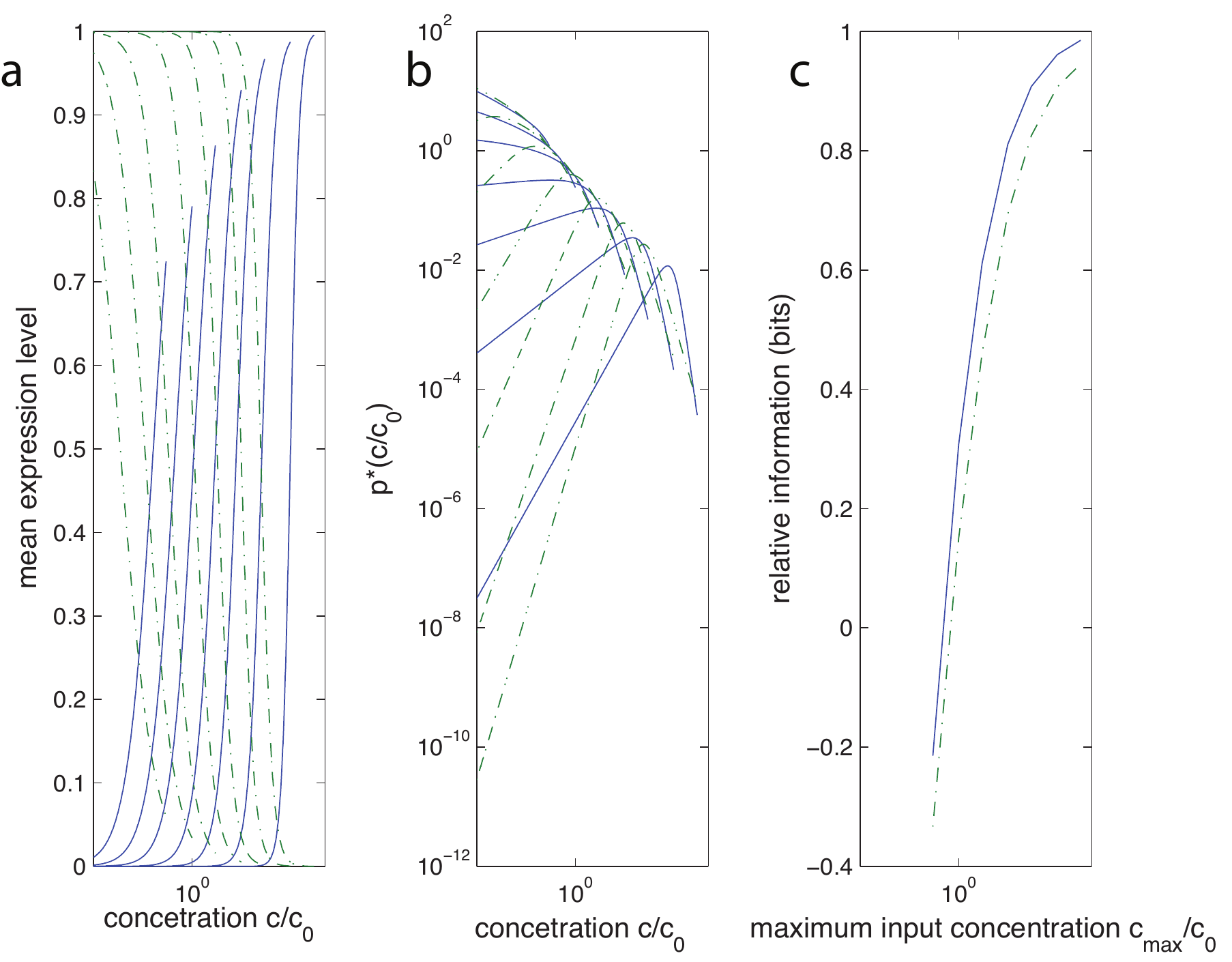}
\caption{(Color online) The optimal solutions for one gene controlled by one transcription factor.   The optimization of information transmission in the small noise limit depends on only one parameter, which we take here as the maximum concentration of the input molecules, measured in units determined by the noise itself [$c_0$ from Eq (\ref{c0})]. Panel A shows the optimal input/output relations with $c_{\rm max}/c_0 = 0.3, \, 1,\, 3,\, 10,\, 30,\, 100,\, 300$; activators shown in blue (solid line), repressors in green (dashed line).   Although the input/output relation is defined for all $c$, we show here only the part of the dynamic range that is accessed when $0 < c < c_{\rm max}$.  
Panel B shows the optimal distributions, $P_{TF}^*(c)$, for each of these solutions. 
Panel C plots $\log_2 \tilde Z_1$ for these optimal solutions as a function of $c_{\rm max}/c_0$.  Up to an additive constant, this is the optimal information capacity, in bits.
\label{i-o2}}
\end{figure}

To proceed, we choose $c_0$ as the unit of concentration, so that
\begin{eqnarray}
Z_1 &=& \left[ {{N_{\rm max}}\over {2\pi e}}\right]^{1/2} \tilde Z_1\\
\tilde Z_1 (K/c_0,n; C) &=&
\int_0^{C} dx\, \left[
{{(d\bar g(x)/dx)^2}
\over{\bar g(x) +  x (d\bar g(x)/dx)^2}
}
\right]^{1/2} ,\nonumber\\
&&\label{normZ1}
\end{eqnarray}
where $C=c_{\rm max}/c_0$ and
\begin{equation}
\bar g(x) = {{x^n}\over{(K/c_0)^n + x^n}}
\label{actHillnorm}
\end{equation}
in the case of an activator.  It now is straightforward to explore, numerically, the function $\tilde Z_1$.  An example, with $c_{\rm max}/c_0 =1$, is shown in Fig \ref{Zmap1}.

We see that, with $c_{\rm max} = c_0$, there is a well defined but broad optimum of the information transmission as a function of the 
parameters $K$ and $n$ describing the input/output relation.  Maximum information transmission occurs at  modest levels  of cooperativity ($n\approx 2$) and with the midpoint of the input/output relation  near the midpoint of the available dynamic range of input concentrations ($K\approx c_{\rm max}/2$).

Optimal solutions for activators and repressors have qualitatively similar behaviors, with the optimal parameters $K_{\rm opt}$ and $n_{\rm opt}$ both increasing as $c_{\rm max}$ increases (Fig \ref{i-o2}a).  Interestingly, at the same value of $c_{\rm max}$, the optimal repressors make fuller use of the dynamic range of outputs.  The information capacity itself, however, is almost identical for activators and repressors, across a wide range of $c_{\rm max}$ (Fig \ref{i-o2}c).   This is important, because it shows that our optimization problem, even in this simplest form, can have multiple nearly degenerate solutions.  We also see that increases of $c_{\rm max}$ far beyond $c_0$ produce a rapidly saturating information capacity, as expected from Eq (\ref{smallc0}). Therefore, although increasing the dynamic range always results in an increase of capacity, the advantage in terms of information capacity gained by the cell being able to use input concentration regimes much larger than $c_0$ is quite small.

\subsection{Some analytic results}
\label{sec_aa}

Although the numerical results are straightforward, we would like to have some intuition about these optimal solutions from analytic approximations.  Our basic problem is to do the integral defining $\tilde Z_1$, in Eq (\ref{normZ1}).  We know that this integral becomes simple in the limit that $C$ is either large or small, so let's start by trying to generate an approximation that will be valid at large $C$. 

At large $C$,  the concentration of input molecules can become large, so we expect that the `output noise,' $\sigma^2 \propto \bar g$, will be dominant.  This suggests that we write
\begin{eqnarray}
\tilde Z_1 &\equiv& \int_0^{C} dx\, \left[
{{(d\bar g(x)/dx)^2}
\over{\bar g(x) +  x (d\bar g(x)/dx)^2}
}
\right]^{1/2}\nonumber\\
&\approx& \int_0^{C} dx\, 
{\Bigg |} {{d\bar g(x)}\over {dx}}{\Bigg |} {1\over \sqrt{\bar g(x)}}\nonumber\\
&&\,\,\,\,\,\times
\left[ 1 - {1\over 2} x {1\over {\bar g(x)}} \left( {{d\bar g(x)}\over {dx}} \right)^2 + \cdots \right] .
\end{eqnarray}
To proceed, we note the combination $dx |d\bar g /dx|$, which invites us to convert this into an integral over $\bar g$.  We use the fact that, for activators described by the Hill function in Eq (\ref{actHillnorm}), 
\begin{eqnarray}
x &=& {K\over{c_0}} \left( {{\bar g}\over{1-\bar g}}\right)^{1/n} ,\\
{{d\bar g(x)}\over {dx}}  &=& {n\over x} \bar g (1-\bar g) .
\label{dgdx1}
\end{eqnarray}
Substituting, we find
\begin{eqnarray}
\tilde Z_1 &\approx& \int_0^{\bar g (C)} 
 {{d\bar g}\over  \sqrt{\bar g}}
\left[ 1 - {{c_0n^2}\over {2K}} \bar g^{1-1/n} (1-\bar g)^{2+ 1/n} + \cdots \right] \nonumber\\
&&\\
&=& 2\sqrt{\bar g (C)} \nonumber\\
&&\,\,\,\,\,- {{c_0n^2}\over {2K}} \int_0^{\bar g(C)} d\bar g \, \bar g^{1/2 - 1/n} (1-\bar g)^{2+1/n} + \cdots .\nonumber\\
&&
\label{step}
\end{eqnarray}
Again, we are interested in large $C$, so we can approximate
$\bar g(C) \approx 1 - (K/c_{\rm max})^n$.  Similarly, the second term in Eq (\ref{step}) can be approximated by letting the upper limit on the integral approach $1$; the difference between $\bar g(C)$ and $1$ generates higher order terms in powers of $1/C$.  Thus we have
\begin{eqnarray}
\tilde Z^{\rm act}_1 &\approx& 2 - \left( {K\over{c_{\rm max}}}\right)^n
- A(n) {{c_0n^2}\over {2K}} + \cdots ;
\label{Zapprox_act}\\
A(n) &=& \int_0^1 dz z^{1/2- 1/n} (1-z)^{2+1/n}  
\\
&=& {{\Gamma(3/2-1/n) \Gamma(3+1/n)}\over{\Gamma(9/2)}} .
\end{eqnarray}

The approximate expression for $\tilde Z_1$ expresses the basic compromise involved in optimizing information transmission.  On the one hand, we would like $K$ to be small so that the output runs through its full dynamic range; correspondingly we want to decrease the term $(K/c_{\rm max})^n$.  On the other hand, we want to move the most sensitive part of the input/output relation to higher concentrations, so that we are less sensitive to the input noise; this corresponds to decreasing the term $\propto c_0/K$.    The optimal compromise is reached at
\begin{equation}
K_{\rm opt}^{\rm  act} \approx c_{\rm max} \left[ {{nA(n) c_0}\over{2c_{\rm max}}}\right]^{1\over{n+1}} .
\label{approxKopt_act}
\end{equation}
Parallel arguments yield, for repressors,
\begin{eqnarray}
\tilde Z^{\rm rep}_1 &\approx& 2 - 2\left(  {K\over{c_{\rm max}}}\right)^n
-B(n) {{c_0n^2}\over {2K}} + \cdots ;
\label{Zapprox_rep}\\
K_{\rm opt}^{\rm  rep} &\approx& c_{\rm max} \left[ {{nB(n) c_0}\over{2c_{\rm max}}}\right]^{2\over{n+2}} ;
\label{approxKopt_rep}\\
B(n) &=&  \int_0^1 dz\, z^{1/2+1/n} (1-z)^{2-1/n} 
\\
&=& {{\Gamma(3/2+1/n) \Gamma(3-1/n)}\over{\Gamma(9/2)}} .
\end{eqnarray}

The first thing we notice about our approximate results is that the optimal values of $K$ are almost proportional to $c_{\rm max}$, as one might expect, but not quite---the growth of $K$ with $c_{\rm max}$ is slightly sublinear.  Also, one might have expected that $K$ would be chosen to divide the available dynamic range into roughly equal `on' and `off' regions, which should maximize the entropy of the output and hence increase the capacity; to achieve this requires $K_{\rm opt}/c_{\rm max} \approx 1/2$.  In fact we see that the ratio $K_{\rm opt}/c_{\rm max}$ is determined by a combination of terms, and depends in an essential way on the scale of the input noise $c_0$, even though we assume that the maximal concentration is large compared with this scale.

The basic compromise between extending the dynamic range of the outputs and avoiding low input concentrations works differently for activators and repressors.  As a result, the optimal values of $K$ are different in the two cases. From Eq (\ref{normZ1}), it is clear that the symmetry between the two types of regulation is broken by the noise term proportional to $\bar g$. Unless the optimal Hill coefficient for repressors were very much smaller than for activators (and it is not), Eqs (\ref{approxKopt_act}) and (\ref{approxKopt_rep}) predict that $K_{\rm opt}^{\rm rep}$ will be smaller than $K_{\rm opt}^{\rm act}$, in agreement with the numerical results in  Fig \ref{i-o2}.

\begin{figure}
\includegraphics[width = \linewidth]{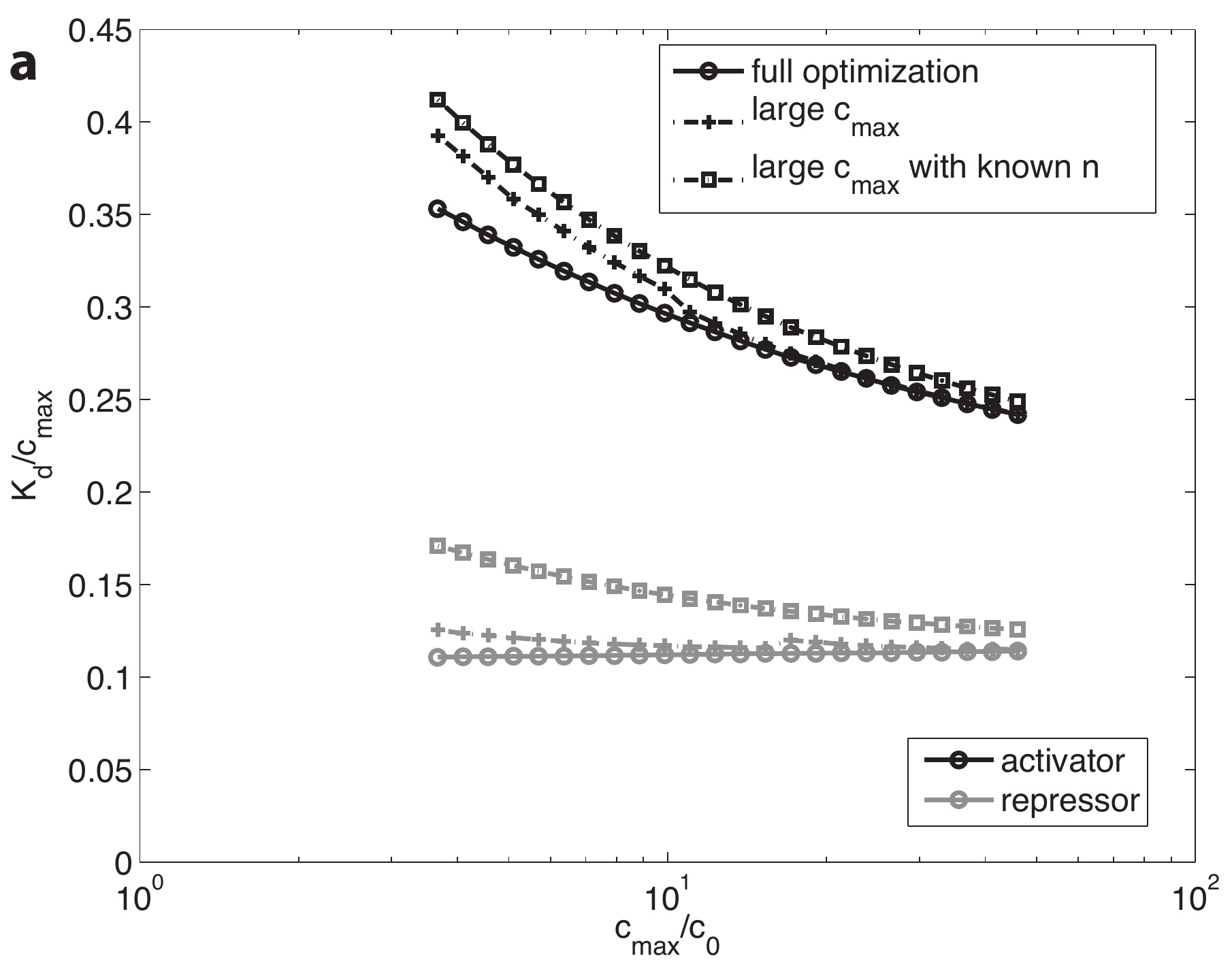}
\includegraphics[width = \linewidth]{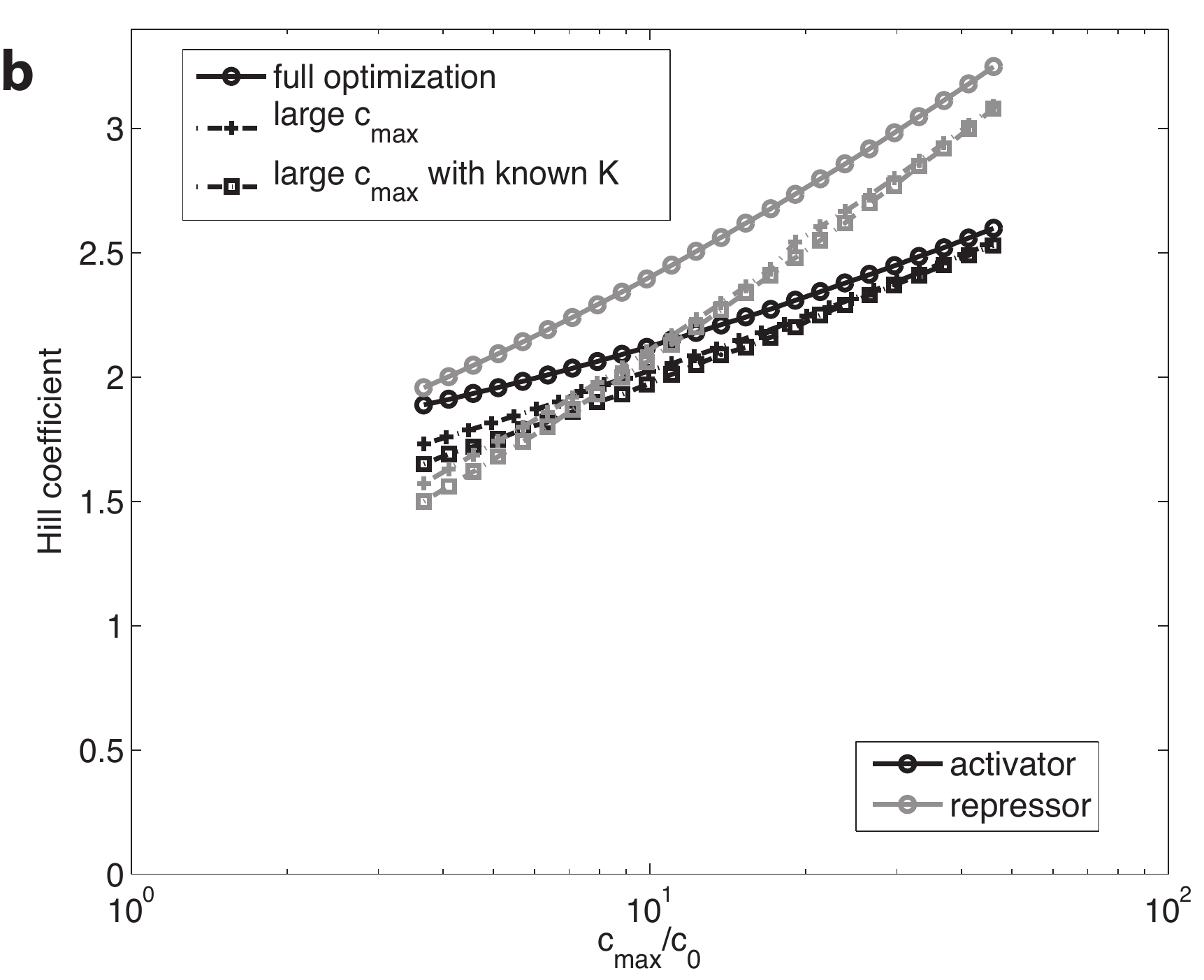}
\caption{Approximate results for the optimal values of $K$ (A)  and $n$ (B) compared with exact numerical results for activators (black lines) and repressors (gray lines).
As explained in the text, we can use our analytic approximations to determine, for example, the optimal $K$ assuming $n$ is known (large $c_{\rm max}$ with known n results), or we can simultaneously optimize both parameters (large $c_{\rm max}$ results); results are shown for both calculations.
\label{Kapprox}}
\end{figure}

To test these analytic approximations, we can compare the predicted values of $K_{\rm opt}$ with those found numerically.  There is a slight subtlety, since our analytic results for $K_{\rm opt}$ depend on the Hill coefficient $n$.  We can take this coefficient as known from the numerical optimization, or we can use the approximations to $\tilde Z_1$ [as in Eq (\ref{Zapprox_act})] to simultaneously optimize for $K$ and $n$.  In contrast to the optimization of $K$, however, there is no simple formula for $n_{\rm opt}$, even in our approximation at large $c_{\rm max}$. 

Results for the approximate vs. numerically exact optimal $K$ are shown in Fig \ref{Kapprox}.  As it should, the approximation approaches the exact answer as $c_{\rm max}$ become large.  In fact, the approximation is quite good even at $c_{\rm max}/c_0 \sim 10$, and for activators the error in $K_{\rm opt}$ is only $\sim 15\%$ at $c_{\rm max}/c_0 \sim 3$.  Across the full range of $c_{\rm max}/c_0 > 1$, the analytic approximation captures the basic trends:  $K_{\rm opt}/c_{\rm max}$ is a slowly decreasing function of $c_{\rm max}/c_0$, $K_{\rm opt}^{\rm act}$ is larger than $K_{\rm opt}^{\rm rep}$ by roughly a factor of $2$, and for both activators and repressors we have $K_{\rm opt}$ noticeably smaller than $c_{\rm max}/2$.  Similarly good results are obtained for the approximate predictions of the optimal Hill coefficient $n$, as shown in Fig \ref{Kapprox}B.

 {\color{black}As noted above, the large $c_{\rm max}$ approximation makes clear that optimizing information transmission is a compromise between using the full dynamic range of outputs and avoiding expression levels associated with large noise at low concentration of the input. The constraint of using the full dynamic range pushes the optimal $K$  downward; this constraint is stronger for repressors [compare the second terms of Eqs (\ref{approxKopt_act}) \&  (\ref{approxKopt_rep})], causing the optimal $K$s of repressors to be smaller than those of the activators. On the other hand, avoiding input noise pushes the most sensitive part of the expression profile toward high concentrations, favoring large $K$. The fact that this approximation captures the basic structure of the numerical solution to the optimization problem encourages us to think that this intuitive compromise is the essence of the problem. It is also worth noting that as $c_{\rm max}$ increases, activators increase their output range, hence gaining capacity. On the other hand, the output of the repressed systems is small for large $c_{\rm max}$ and the output noise thus is large, limiting the increase in capacity compared to the activated genes, as is seen in Fig \ref{i-o2}c.}

In the case of small $c_{\rm max}$ it is harder to obtain detailed expressions for $K$, however we can still gain insight from the expression for the capacity in this limit.  To obtain the large $c_{\rm max}$ limit we assumed that $\bar g \gg x (d\bar g/dx)^2$ in the denominator of the integrand which defines $Z_1$; to obtain the small $c_{\rm max}$ limit we make the opposite assumption:
\begin{eqnarray}
\tilde Z_1 &\approx& \int_0^{C}
dx\, \left[
{{(d\bar g(x)/dx)^2}
\over{\bar g(x) +  x (d\bar g(x)/dx)^2}
}
\right]^{1/2}\nonumber\\
&=& \int_0^{C} dx\,
\frac{1}{\sqrt{x}}  \left[
{{1}
\over{1+\bar g(x)/(x(d\bar g(x)/dx)^2) }
}
\right]^{1/2}\nonumber\\
&\approx& \int_0^{C} dx\,
\frac{1}{\sqrt{x}} \left[1-\frac{x}{2n^2} \frac{1}{g(1-g)^2}+ \cdots \right] ,
\label{step0}
\end{eqnarray}
where in the last step we use the relation in Eq (\ref{dgdx1}).   We see that, if $g$ approaches one, the first correction term will diverge.  This  allows us to predict the essential feature of the optimal solutions at small $c_{\rm max}$, namely that they do not access the full dynamic range of outputs.

\subsection{Constraining means}
\label{sec_conmean}

\begin{figure*}
\includegraphics[width = 0.4\linewidth]{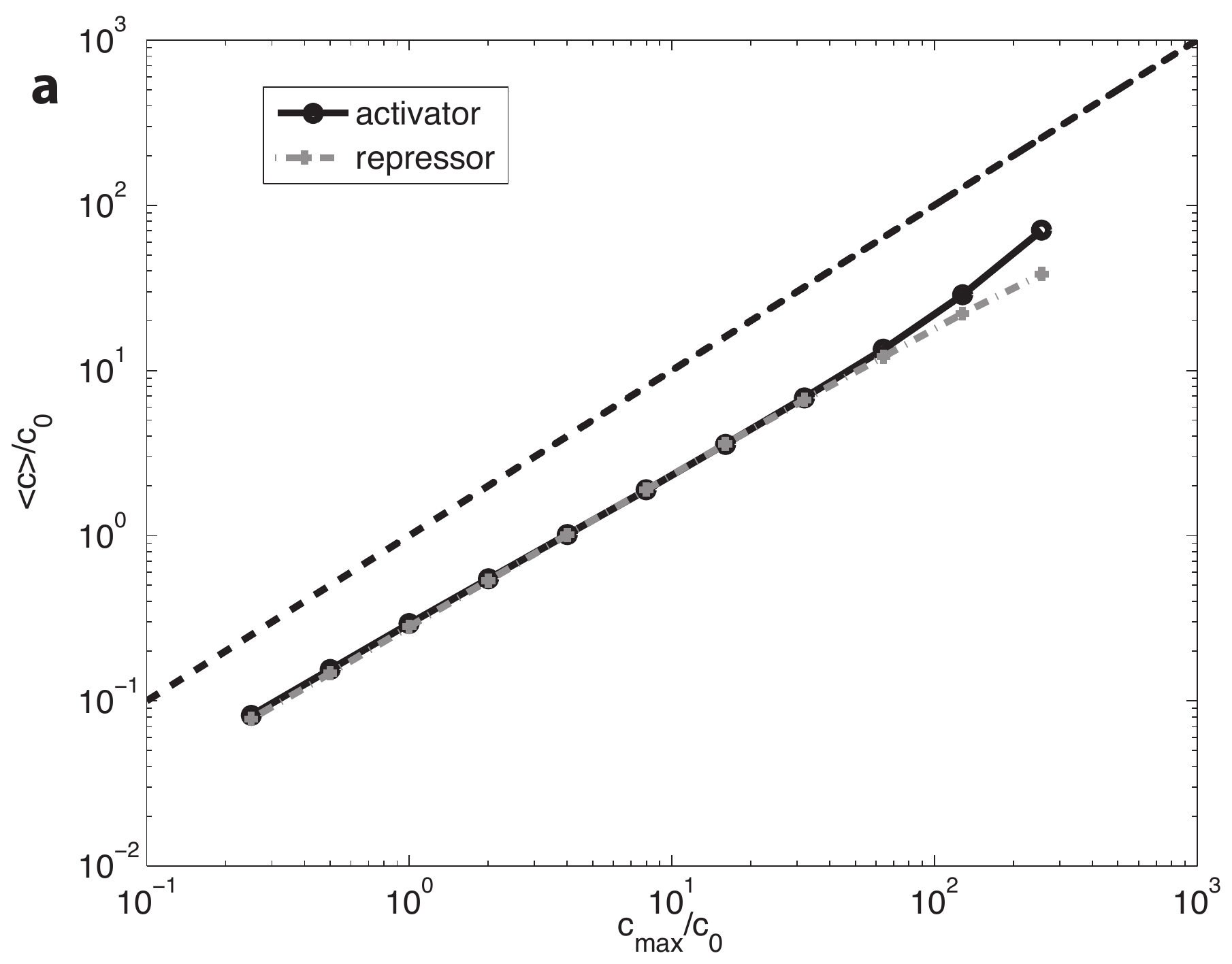}
\includegraphics[width = 0.4\linewidth]{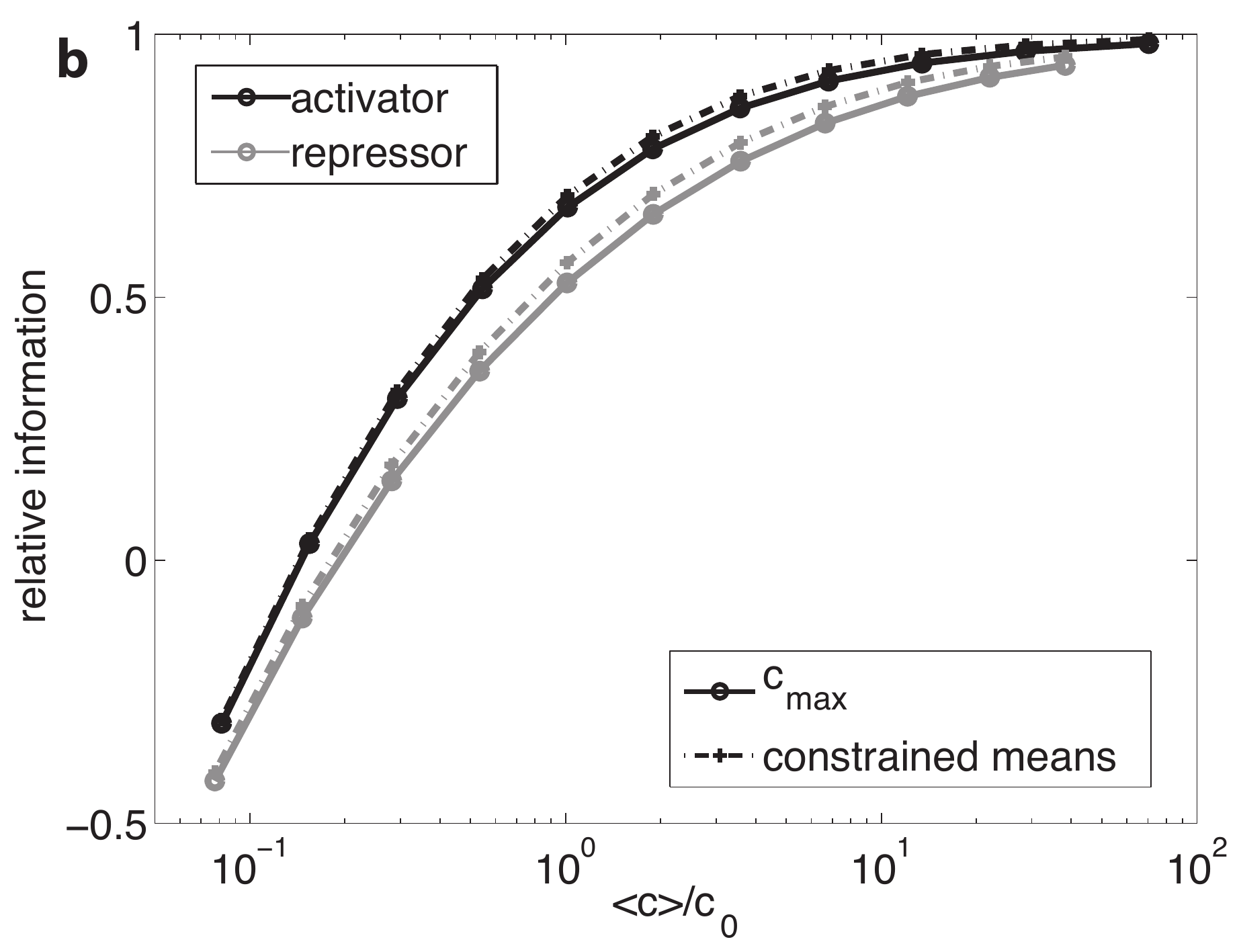}
\includegraphics[width = 0.4\linewidth]{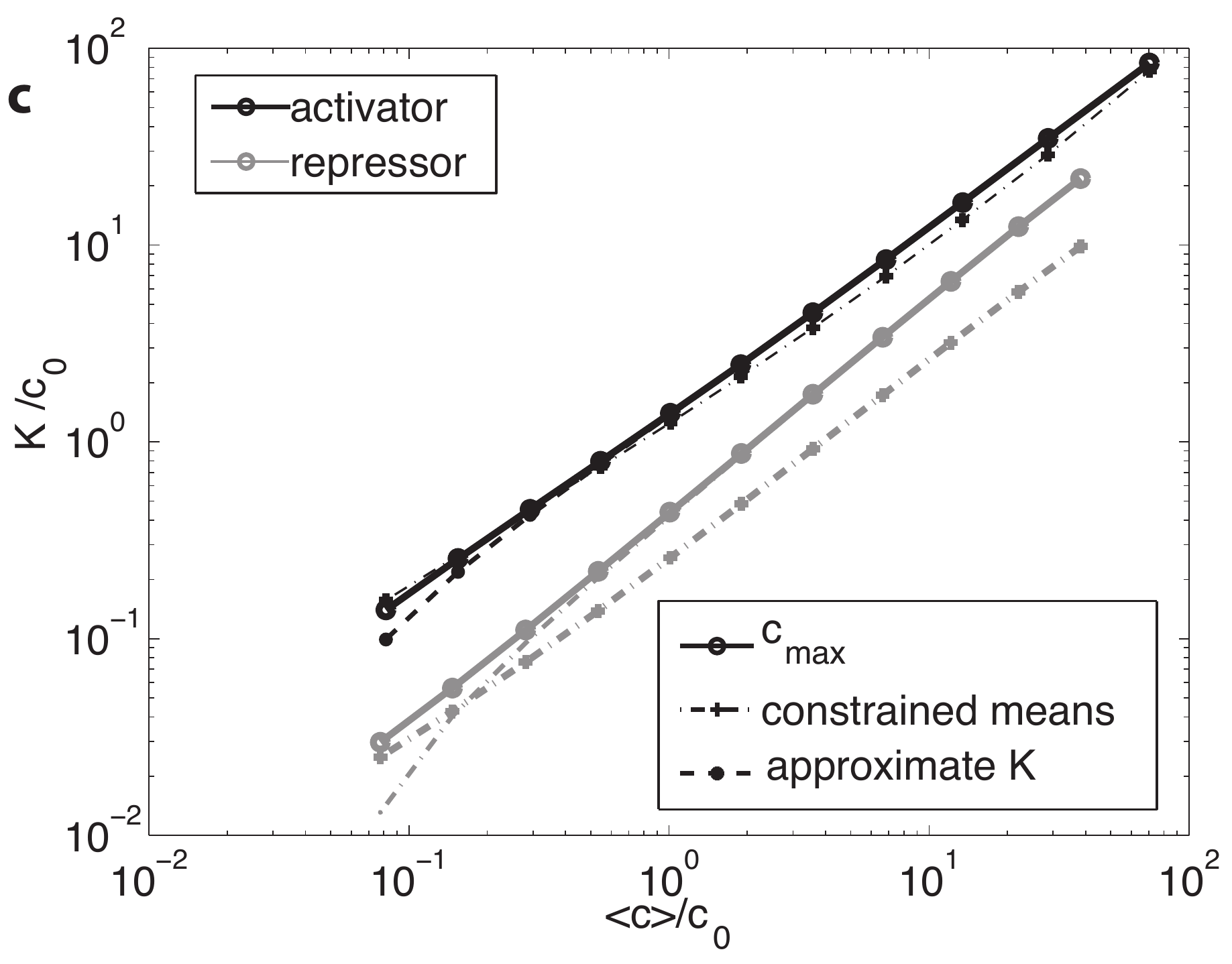}
\includegraphics[width = 0.4\linewidth]{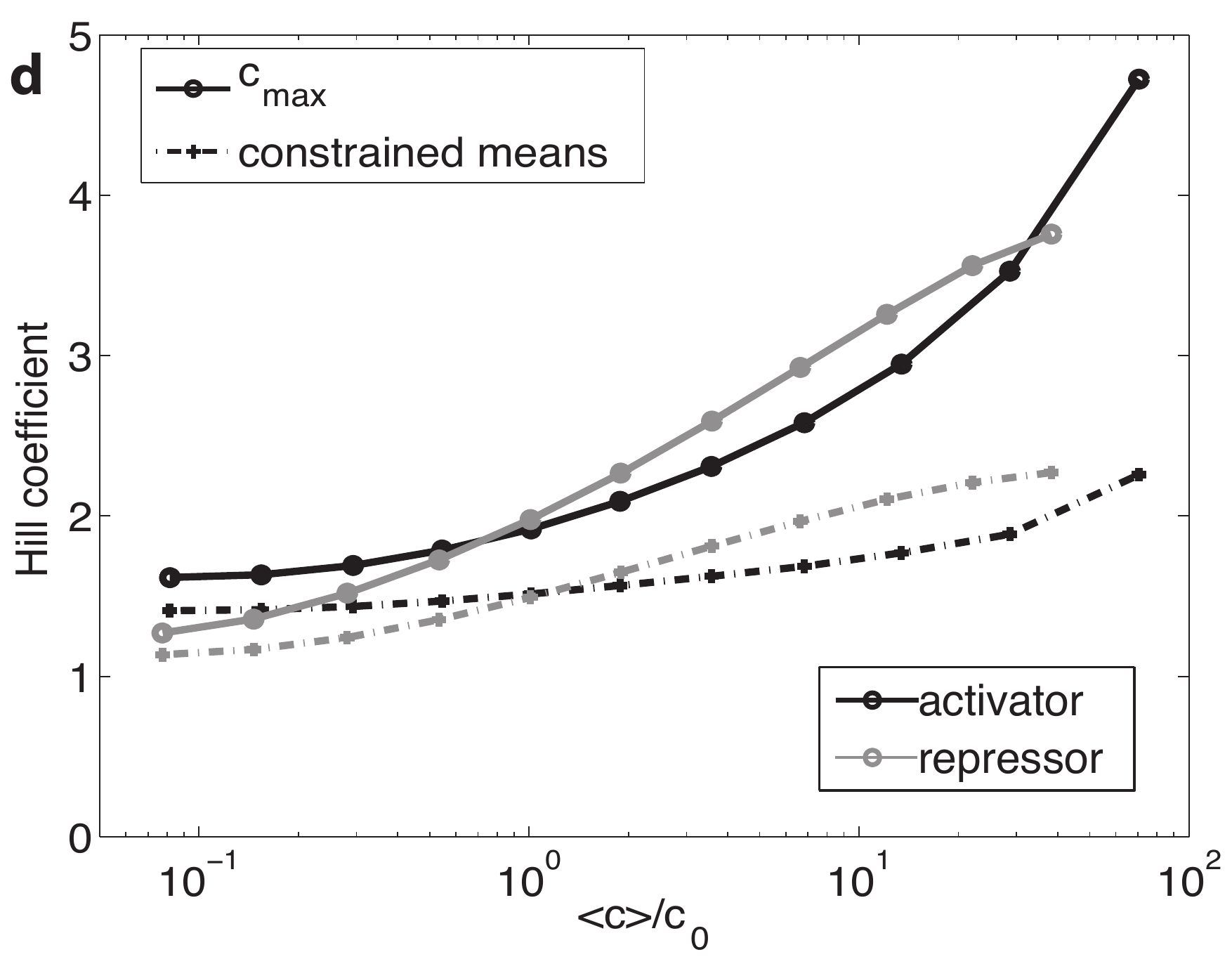}
\caption{A: Mean concentration of the transcription factor when we optimize information transmission subject to a constraint on the maximum concentration.  Results are shown for one input and one output, both for activators and repressors. The dashed black line shows equality. B-D: Comparing two formulations of the optimization problem for activators (black lines) and repressors (gray lines) calculated with a finite dynamic range ($c_{\rm max}$ - circles and solid lines) and constrained means (crosses and dashed lines). The panels show the relative information in panel B, the optimal value of $K$ in panel C, the optimal value of the Hill coefficient in panel D. In panel C, approximate results for $K$ are shown as a function of $\langle c\rangle$, from    Eq's (\ref{actKvcc})   and  (\ref{repKvcc}).
\label{compare}}
\end{figure*}

Here we would like to solve the same optimization problem by constraining the mean concentrations, rather than imposing a hard constraint on the maximal concentrations; as noted above we can also think of this problem as maximizing information subject to a fixed cost per molecule. To compare results in a meaningful way, we should know how the mean concentration varies as a function of $c_{\rm max}$ when we solve the problem with constrained maxima, and this is shown in Fig \ref{compare}A.  An interesting feature of these results is that mean concentrations are much less than half of the maximal concentration. Also, the mean input concentrations for activator and repressor systems are similar, despite different  values of the optimal $K$. This result shows that for a given dynamic range defined by $c_{\rm max}$, there is an optimal mean input concentration, which is independent of whether the  input/ouput relation is up or down regulating.

Equation (\ref{Z2def}) shows us how to compute the partition function $Z_2$ for the case where we constrain the mean concentration of transcription factors, and Eq (\ref{I2def}) relates this to the information capacity $I_2$.  Substituting our explicit expressions for the noise in the case of one input and one output, we have
\begin{eqnarray}
Z_2 &=& \left[ {{N_{\rm max}}\over{2\pi e}}\right]^{1/2} \tilde Z_2\\
\tilde Z_2 &=& \int_0^\infty dc
\left[
{{(d\bar g(c)/dc)^2}
\over{\bar g(c) +  c c_0 (d\bar g(c)/dc)^2}
}
\right]^{1/2} e^{-\alpha c} .
\end{eqnarray}

As before, we choose Hill functions for $\bar g(c)$, and maximize $I_2$ with respect to the parameters $K$ and $n$.  This defines a family of optimal solutions parameterized by 
 the Lagrange multiplier $\alpha$, and we can tune this parameter to match the mean concentration $\langle c \rangle$. Using the calibration in Fig \ref{compare}A, we can compare these results with those obtained by optimizing with a fixed maximum concentration.  Results are shown in Fig \ref{compare}b--d.

The most important conclusion from Fig \ref{compare} is that constraining mean concentrations and constraining maximal concentrations give---for this simple problem of one input and one output---essentially the same answer.  The values of the optimal $K$s are almost identical (Fig \ref{compare}C), as are the actual number of bits that can be transmitted (Fig \ref{compare}B).  The only systematic difference is in the Hill coefficient $n$, where having  a fixed maximal concentration drives the optimization toward slightly larger values of $n$ (Fig \ref{compare}D), so that more of the dynamic range of outputs is accessed before the system runs up against the hard limit at $c=c_{\rm max}$.

It is  interesting that the optimal value of $K$ is more nearly a linear function of $\langle c \rangle$ than of $c_{\rm max}$, as we see in Fig \ref{compare}C.    To understand this, we follow
the steps in Section \ref{sec_aa}, expanding the expression for $\langle c \rangle$ in the same approximation that we used for large $c_{\rm max}$:
\begin{eqnarray}
\langle c \rangle &=& {\int_0^{C} dc\ c \left[
{(d\bar g(c)/dc)^2}
\over{\bar g(c) +  c c_0 (d\bar g(c)/dc)^2}
\right]^{1/2} }
\over{\int_0^{C} dc 
\left[ {(d\bar g(c)/dc)^2}
\over{\bar g(c) +  c c_0 (d\bar g(c)/dc)^2}
\right]^{1/2}}
\nonumber\\
& \approx &{ \int_{\bar g(0)}^{\bar g (C)}
 {d\bar g}\ {{c}\over  \sqrt{\bar g}} - \frac{1}{2}  \int_{\bar g(0)}^{\bar g (C)} 
 {{d\bar g}\ n^2 {\bar g}^{\frac{1}{2}} (1-\bar g)^2 }} \over{ \int_{\bar g(0)}^{\bar g (C)} 
 {d\bar g} {1\over  \sqrt{\bar g}}  - \frac{1}{2}  \int_{\bar g(0)}^{\bar g (C)} 
 {{d\bar g}\ n^2 c{\bar g}^{\frac{1}{2}} (1-\bar g)^2 }}
\end{eqnarray}

In the case of an activator, $c = {K/{c_0}} \left( {{\bar g}/{(1-\bar g)}}\right)^{1/n} $, and the leading terms become:
\begin{eqnarray}
\langle c\rangle &=& \frac{\int_{\bar g(0)}^{\bar g (C)}d\bar g\ {\bar g}^{\frac{1}{n}-\frac{1}{2}} (1-\bar g)^{-1/n}}{\int_{\bar g(0)}^{\bar g (C)}d\bar g\   {\bar g}^{-\frac{1}{2}}} \nonumber\\
&& \times \left[ K + \frac{n^2}{2} {{\int_{\bar g(0)}^{\bar g (C)}d\bar g\   {\bar g}^{\frac{1}{2}-\frac{1}{n}} (1-\bar g)^{2+\frac{1}{n}}}\over{\int_{\bar g(0)}^{\bar g (C)}  {\bar g}^{-\frac{1}{2}}}}
+...\right] \nonumber\\
& &
- \frac{n^2}{2} {{\int_{\bar g(0)}^{\bar g (C)}d\bar g\   {\bar g}^{\frac{1}{2}} (1-\bar g)^{2}}\over{\int_{\bar g(0)}^{\bar g (C)}d\bar g\   {\bar g}^{-\frac{1}{2}}}}.
\label{actKvcc}
\end{eqnarray}
To get some intuition for the numerical values of these terms we will assume the integral covers the whole expression range $\bar g \in [0, 1]$, and $n=3$. Then this expression simplifies to
\begin{equation}
\label{roughact}
\langle c\rangle \approx 0.86 K+0.52 ,
\end{equation}
so we understand how this simple result emerges, at least asymptotically at large $c_{\rm max}$.

In the case of repressors the leading terms are:
\begin{eqnarray}
\langle c\rangle &=& \frac{\int_{\bar g(0)}^{\bar g (C)}d\bar g\  {\bar g}^{-\frac{1}{n}-\frac{1}{2}} (1-\bar g)^{1/n}}{\int_{\bar g(0)}^{\bar g (C)}d\bar g\   {\bar g}^{-\frac{1}{2}}} \nonumber\\
&& \times \left[ K + \frac{n^2}{2} {{\int_{\bar g(0)}^{\bar g (C)}d\bar g\   {\bar g}^{\frac{1}{2}+\frac{1}{n}} (1-\bar g)^{2-\frac{1}{n}}}\over{\int_{\bar g(0)}^{\bar g (C)}  {\bar g}^{-\frac{1}{2}}}}
+...\right] \nonumber\\
& &
- \frac{n^2}{2} {{\int_{\bar g(0)}^{\bar g (C)}d\bar g\   {\bar g}^{\frac{1}{2}} (1-\bar g)^{2}}\over{\int_{\bar g(0)}^{\bar g (C)}d\bar g\   {\bar g}^{-\frac{1}{2}}}}.
\label{repKvcc}
\end{eqnarray}
As in the case of the activator, making the rough approximation that $n=3$ and $\bar g \in [0, 1]$ allow us to get some intuition for this large $c_{\rm max}$ result:
\begin{eqnarray}
\label{roughrep}
\langle c\rangle &\approx&2.8 K+1.19 ,
\end{eqnarray}
These extremely crude estimates do  predict the basic linear trends in   Fig \ref{compare}C, including the fact that for a given value of the mean concentration, the repressor has a smaller $K$ than the activator. 

Before leaving this section, we should return to the question of constraining mean outputs, as well as mean inputs.  We have measured the input concentration in absolute units (or relative to the physical scale $c_0$), so when we constrain the mean input we really are asking that the system use only a fixed mean number of molecules.  In contrast, we have measured outputs in relative units, so that the maximum of $\bar g(c)$ is one.  If we want to constrain the mean number of output molecules, we need to fix not $\langle g\rangle$, but rather $N_{\rm max}\langle g\rangle$, since the factor of $N_{\rm max}$ brings us back to counting the molecules in absolute terms \cite{remark2}.  Thus, exploring constrained mean output requires us to view $N_{\rm max}$ (and hence the scale $c_0$) as an extra adjustable parameter. 

{\color{black} By itself, adding $N_{\rm max}$ as an additional optimization parameter makes our simple problem more complicated, but does not seem to add much insight.} In principle it would allow us to discuss the relative information gain on adding extra  input vs output molecules, with the idea that we might find optimal information transmission subject to some net resource constraint; for initial results in this direction see Ref \cite{tkacik+al_08b}.  In networks with feedback, the target genes also act as transcription factors, and these tradeoffs should be more interesting.  We will return to this problem in subsequent papers.

\section{Multiple outputs}

When the single transcription factor at the input of our model system has multiple independent target genes, and we constrain the maximal concentrations, the general form of the information capacity in the small noise limit is given by Eq (\ref{Z1_multiple}),
\begin{widetext}
\begin{eqnarray}
Z_1 &=& \int_0^{c_{\rm max}} dc\, \left[{1\over{2\pi e}} \sum_{{\rm i}=1}^M  {1\over {\sigma_{\rm i}^2 (c)}}
\left({{d\bar g_{\rm i}(c)}\over{dc}} \right)^2
\right]^{1/2} \nonumber\\
&=& \left[ {{N_{\rm max}}\over{2\pi e}}\right]^{1/2}
\int_0^{c_{\rm max}} dc\, 
\left[
\sum_{{\rm i} = 1}^M
{{(d\bar g_{\rm i}(c)/dc)^2}\over
{\bar g_{\rm i}(c) + c c_0 (d\bar g_{\rm i}(c)/dc)^2}}
\right]^{1/2} , 
 \label{multipleeq}
\end{eqnarray}
\end{widetext}
where we assume for simplicity  that the basic parameters $N_{\rm max}$ and $D\ell\tau$ are the same for all the target genes
Once again, $c_0 = N_{\rm max}/D\ell\tau$ provides a natural unit of concentration. {\color{black}We limit ourselves to an extended discussion of the case with a hard upper bound, $c_{\rm max}$, to the dynamic range of the input. As in the case of a single output, the calculation with a constrained mean input concentration gives essentially the same results.}

\begin{figure}[b]
\includegraphics[width = \linewidth]{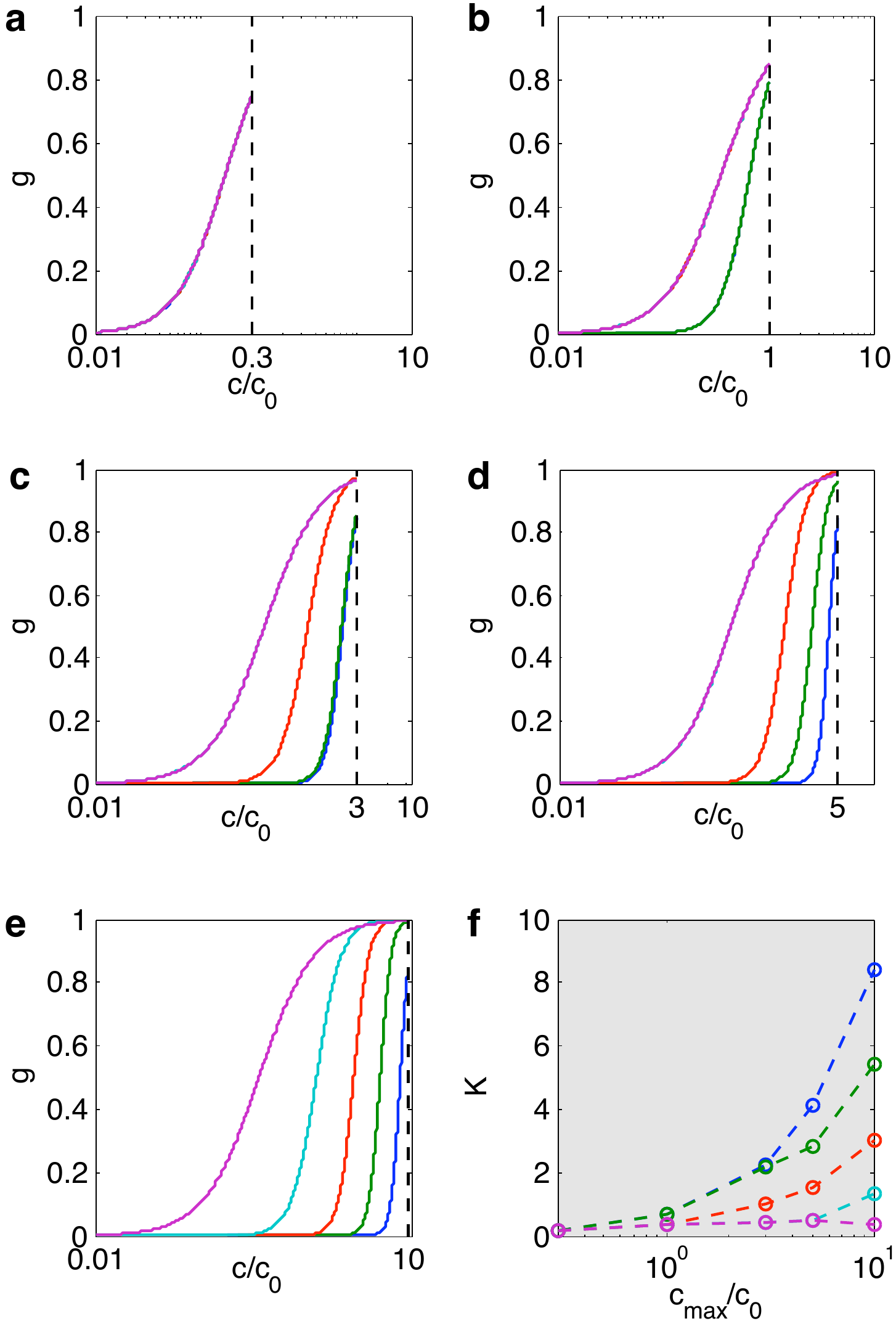}
\caption{(Color online) Optimal input/output relations for the case of five independent target genes, activated by the TF at concentration $c$.  Successive panels (A-E) correspond to different values of the maximal input concentration, as indicated ($C=0.3,1,3,5,10$). Panel F summarizes the optimal values of the $K_{\rm i}$ as a function of $C=c_{\rm max}/c_0$: as $C$ is increased, the $K_i$ of the fully redundant input/output relations for $C=0.3$ bifurcate such that at $C=10$ the genes tile the whole input range.
\label{5genes}}\end{figure}

To get an initial feeling for the structure of the problem, we try the case of five target genes, all of which are activated by the transcription factor. Then 
\begin{equation}
\bar g_{\rm i} (c) = {{c^{n_{\rm i}}}\over {c^{n_{\rm i}} + K_{\rm i}^{n_{\rm i}}}} ,
\end{equation}
and we can search numerically for the optimal settings of all the parameters $\{K_{\rm i}, n_{\rm i}\}$.  Results are shown in Fig \ref{5genes}.  A striking feature of the problem is that, for small values of the maximal concentration $C=c_{\rm max}/c_0$, the optimal solution is actually to have all five target genes be completely redundant, with identical values of $K_{\rm i}$ and $n_{\rm i}$.  As $c_{\rm max}$ increases, this redundancy is lifted, and the optimal solution becomes a sequence of target genes with staggered activation curves, in effect `tiling' the input domain $0 < c < c_{\rm max}$. {\color{black}To interpret these results, we realize that for small maximal concentration the input noise dominates and the optimal strategy for $M$ genes is to `replicate' one well-placed gene $M$-times: having  $M$ independent and redundant readouts (with identical $K$ and $n$) of the input concentration will decrease the noise by a factor of $\sqrt{M}$. However, as the dynamic range increases and output noise has a chance to compete with  the input noise, more information can be transmitted by using $M$ genes to probe the input at different concentrations, thereby creating a cascade of genes that get activated at successively higher and higher input levels. The transition between these two readout strategies is described in more detail below.}

\begin{figure*}
\includegraphics[height=2.1 in]{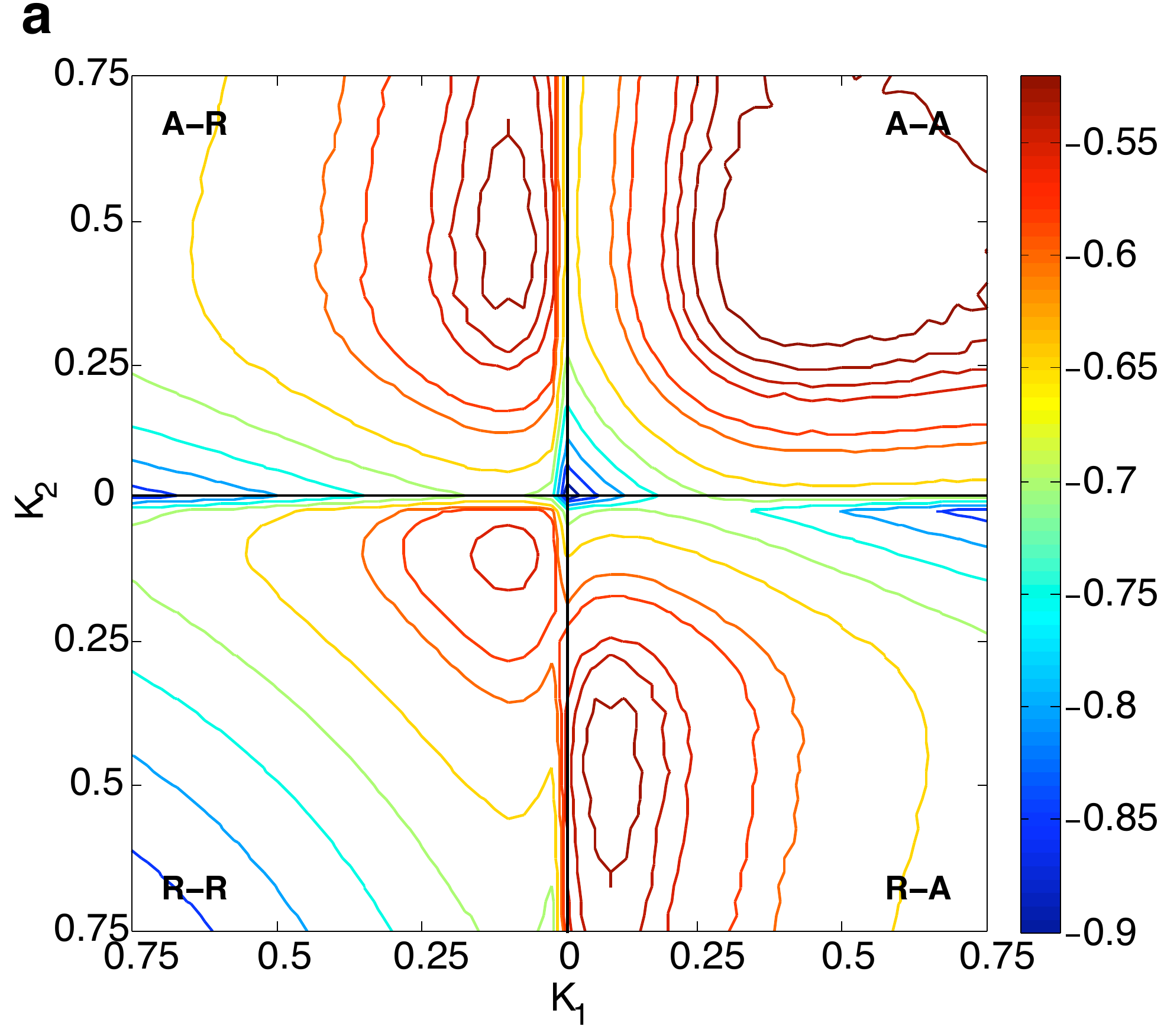}
\includegraphics[height=2.1 in]{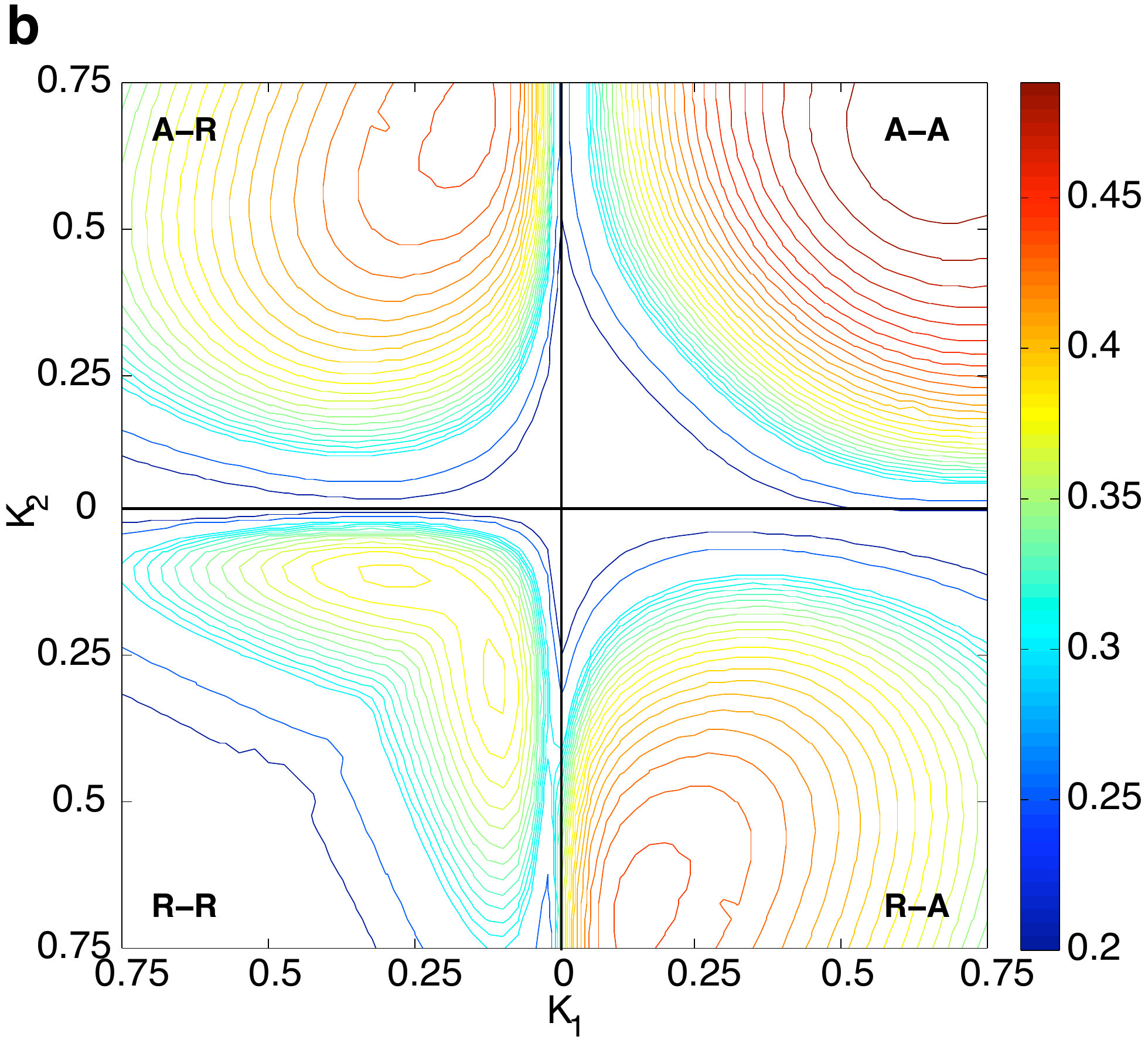}
\includegraphics[height=2.1 in]{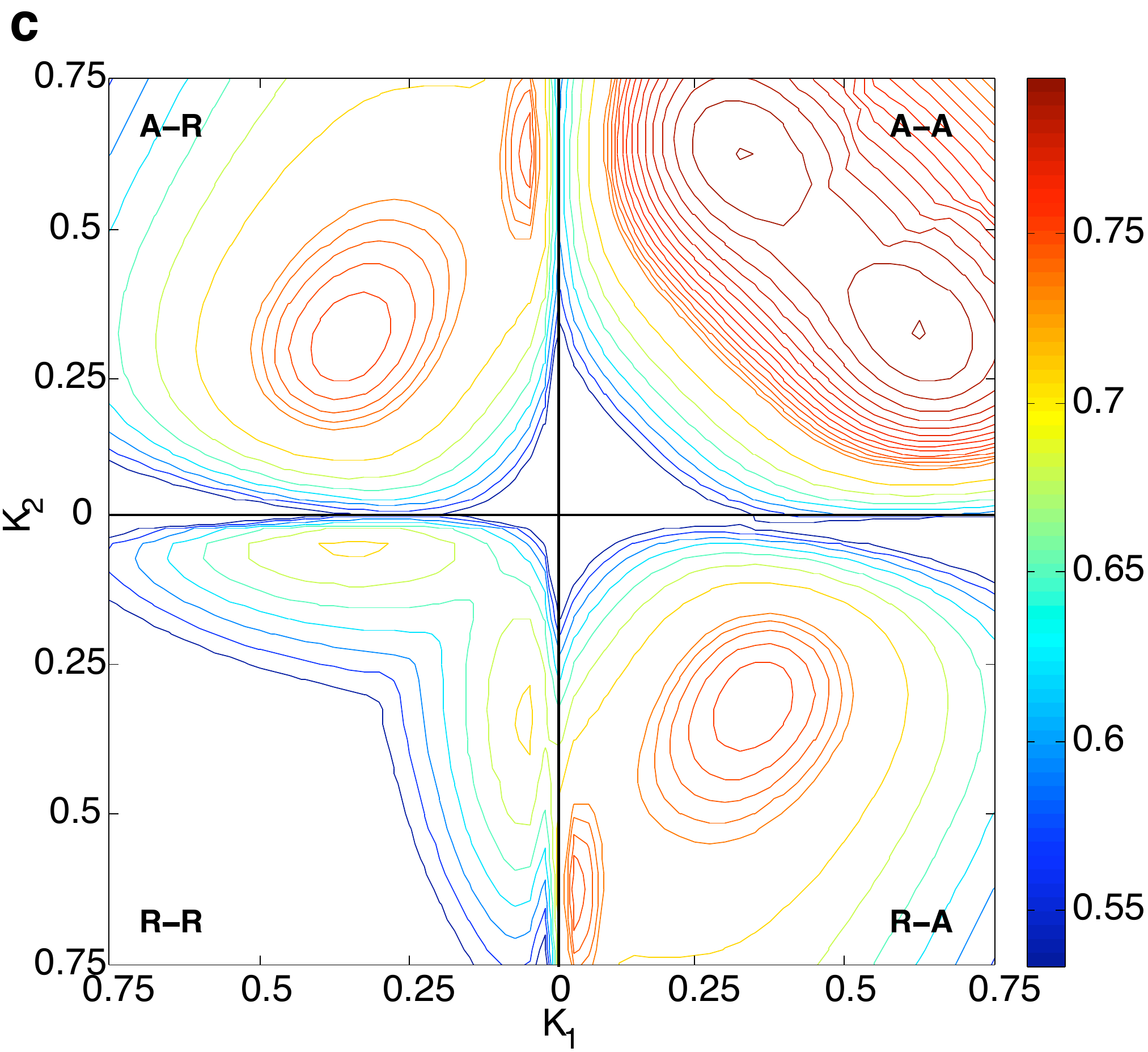}
\caption{(Color online) The case of two target genes. The maps show contour plots of relative information ($\log_2{\tilde{Z}_1}$), as a function of the $K$ values of the two genes: $K_1$ and $K_2$. 
In each map, the upper right quadrant (A-A) contains solutions where both genes are activated by a common TF, in the lower left quadrant (R-R)  both genes are repressed, and the other two quadrants (A-R)  contain an activator-repressor mix. 
The maximal concentration of the input molecules is fixed at $c_{\rm max} /c_0 = 0.1$ in map A, at $0.5$ in map B, and at $1$ in map C.  We see that, for example, only at the highest value of $c_{\rm max}$ does the two activator solution in the upper right quadrant correspond to distinct values of $K_1$ and $K_2$; at lower values of $c_{\rm max}$ the optimum is along the `redundant' line $K_1 = K_2$.  The redundancy is lifted at lower values of $c_{\rm max}$ in the case of repressors, as we see in the lower left quadrants, and  the mixed activator/repressor solutions are always asymmetric.  At large $c_{\rm max}$ we also see that there are two distinct mixed solutions.  
\label{2genemaps}}
\end{figure*}

\begin{figure}[bt]
\includegraphics[width=\linewidth]{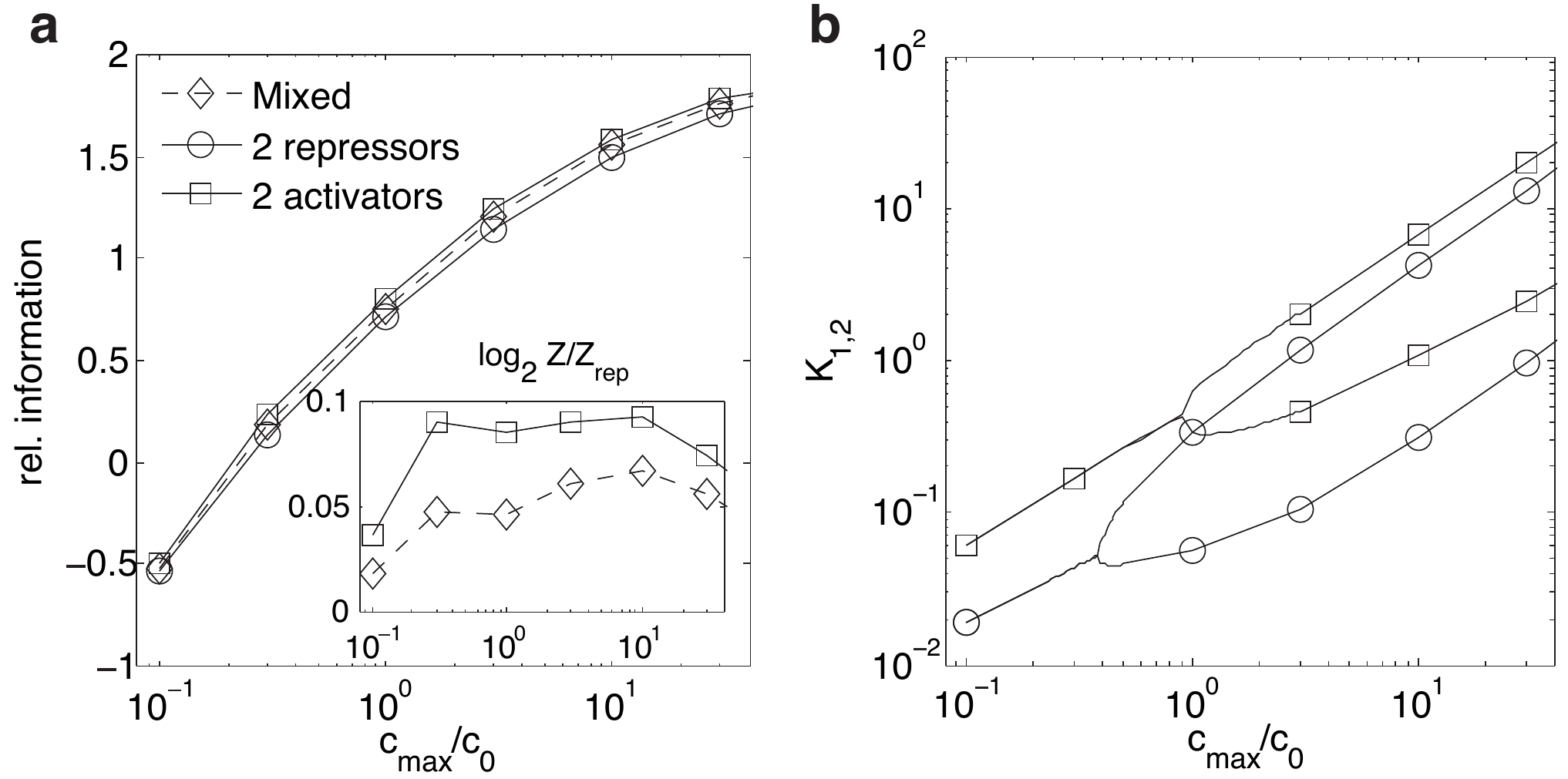}
\caption{The relative information for stable solutions for two genes as a function of $c_{\rm max}$ (panel A). The inset   shows the difference in information transmission for 2 activators and the mixed case, relative to the two repressors. In panel B,  the optimal $K_1$ and $K_2$ are plotted as a function of $c_{\rm max}$ for two activators (squares) and two repressors (circles). The bifurcation in $K$ is a continuous transition that happens at lower $c_{\rm max}$ in the case of two repressors. \label{bif}}
\end{figure}

To look more closely at the structure of the problem, we drop down to consider two target genes.  Then there are three possibilities---two activators (AA), two repressors (RR), and one of each (AR).  For each of these discrete choices, we have to optimize two exponents $(n_1 , n_2 )$ and two half--maximal points $(K_1 , K_2)$.  In Fig \ref{2genemaps} we show how $\tilde{Z}_1$ varies in the $(K_1 , K_2)$ plane, assuming that at every point we choose the optimum exponents; the different quadrants correspond to the different discrete choices of activator and repressor.  The results show clearly how the redundant ($K_1 = K_2 $) solutions at low values of $c_{\rm max}$ bifurcate into asymmetric ($K_1 \neq K_2$) solutions at larger values of $c_{\rm max}$; the critical value of $c_{\rm max}$ is different for activators and repressors.  This bifurcation structure is summarized in Fig \ref{bif}, where we also see that, for each value of $c_{\rm max}$, the three different kinds of solutions (AA, RR and AR) achieve information capacities that differ by less than $0.1$ bits.

The information capacity is an integral of the square root of a sum of terms, one for each target gene [Eq (\ref{multipleeq})].  Thus if we add redundant copies of a single gene, all with the same values of $K$ and $n$, the integral $Z_1$ will scale as $\sqrt{M}$, where $M$ is the number of genes.  In particular, as we go from 1 to 2 target genes, $Z$ would increase by a factor $\sqrt{2}$ and hence the information capacity, $\log_2 Z$, would increase by one half bit; more generally, with $M$ redundant copies, we have $(1/2)\log_2 M$ bits of extra information relative to having just one gene.  On the other hand, if we could arrange for two target genes to make non--overlapping contributions to the integral, then two genes could have a value of $Z$ that is twice as large as for one gene, generating an extra bit rather than an extra half bit.  In fact a full factor of two increase in $Z$ isn't achievable, because once the two target genes are sampling different regions of concentration they are making different tradeoffs between the input and output noise terms; since the one gene had optimized this tradeoff, bifurcating into two distinguishable targets necessarily reduces the contribution from each target.  Indeed, if the maximal concentration is too low then there is no `space' along the $c$ axis to fit two distinct activation (or repression) curves, and this is why low values of $c_{\rm max}$ favor the redundant solutions.

\begin{figure}
\includegraphics[width=\linewidth]{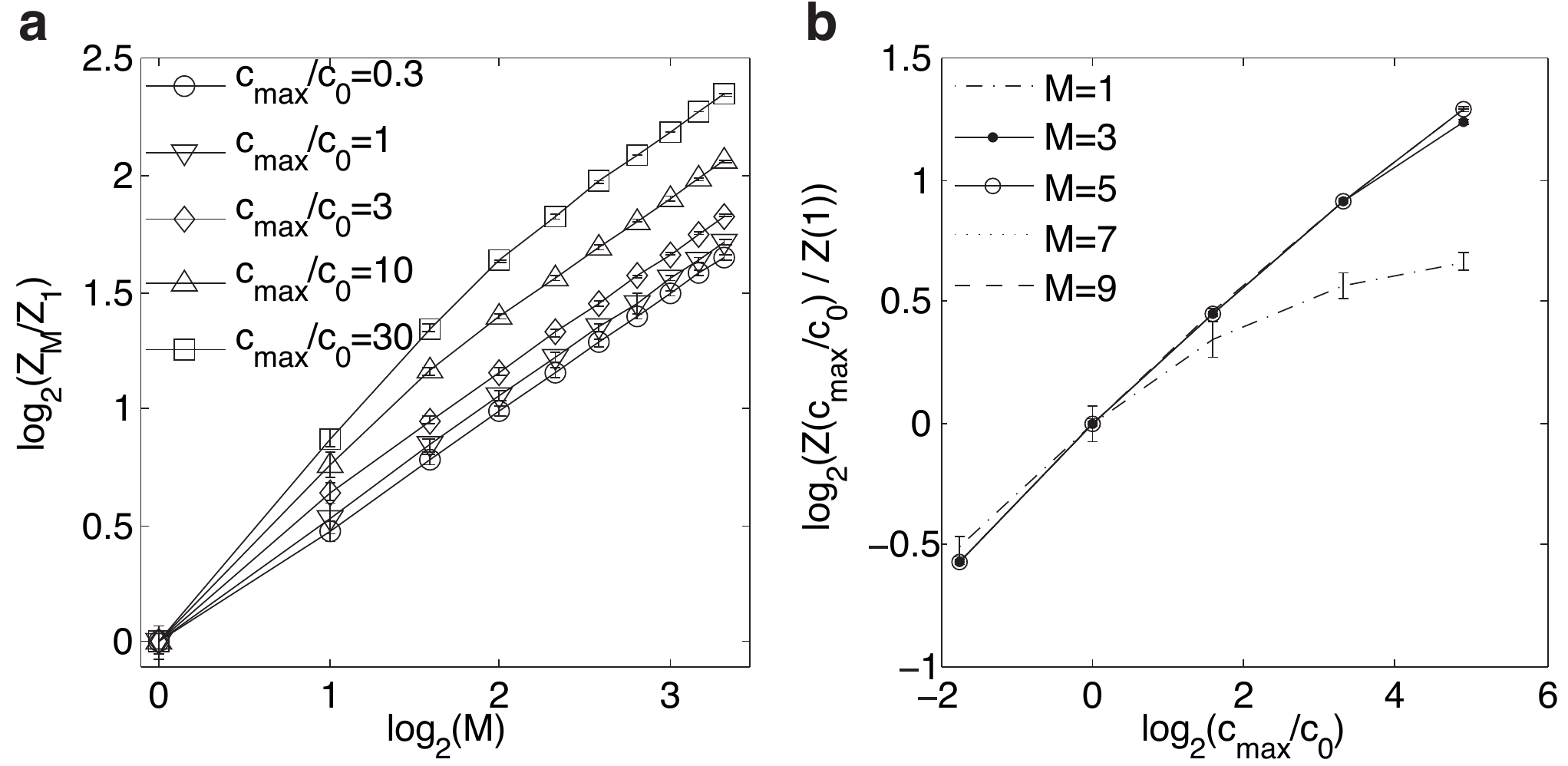}
\caption{The relative information for different values of $c_{\rm max}$ as a function of the number of genes, $M$, shown in panel A. At low $c_{\rm max}$ the genes are redundant and so the capacity grows as $(1/2)\log_2 M$; at high $c_{\rm max}$, the increase in capacity is larger, but bounded from above by one bit. The differences in information for various combinations of activators and repressors are comparable to the size of the plot symbols. In panel B, the relative information for different numbers of genes as a function of $c_{\rm max}$.  At higher $M$, the system can make better use of the input dynamic range.
\label{manygenes}}
\end{figure}

Figure \ref{manygenes}a shows explicitly that when we increase the number of target genes at low values of $c_{\rm max}$, the optimal solution is to use the genes redundantly and hence the gain in information is $(1/2)\log_2 M$.  At larger values of $c_{\rm max}$, going from one target to two targets one can gain more than half a bit, but this gain is bounded by one bit, and indeed over the range of $c_{\rm max}$ that we explore here the full bit is never quite reached.

We can take a different slice through the parameter space of the problem by holding the number of target genes fixed and varying $c_{\rm max}$.  With a single target gene, 
we have seen (Fig \ref{i-o2}) that the information capacity saturates rapidly as $c_{\rm max}$ is increased above $c_0$.    We might expect that, with multiple target genes, it is possible to make better use of the increased dynamic range, and this is what we see in Fig \ref{manygenes}b.

\begin{figure}
\includegraphics[scale=0.5]{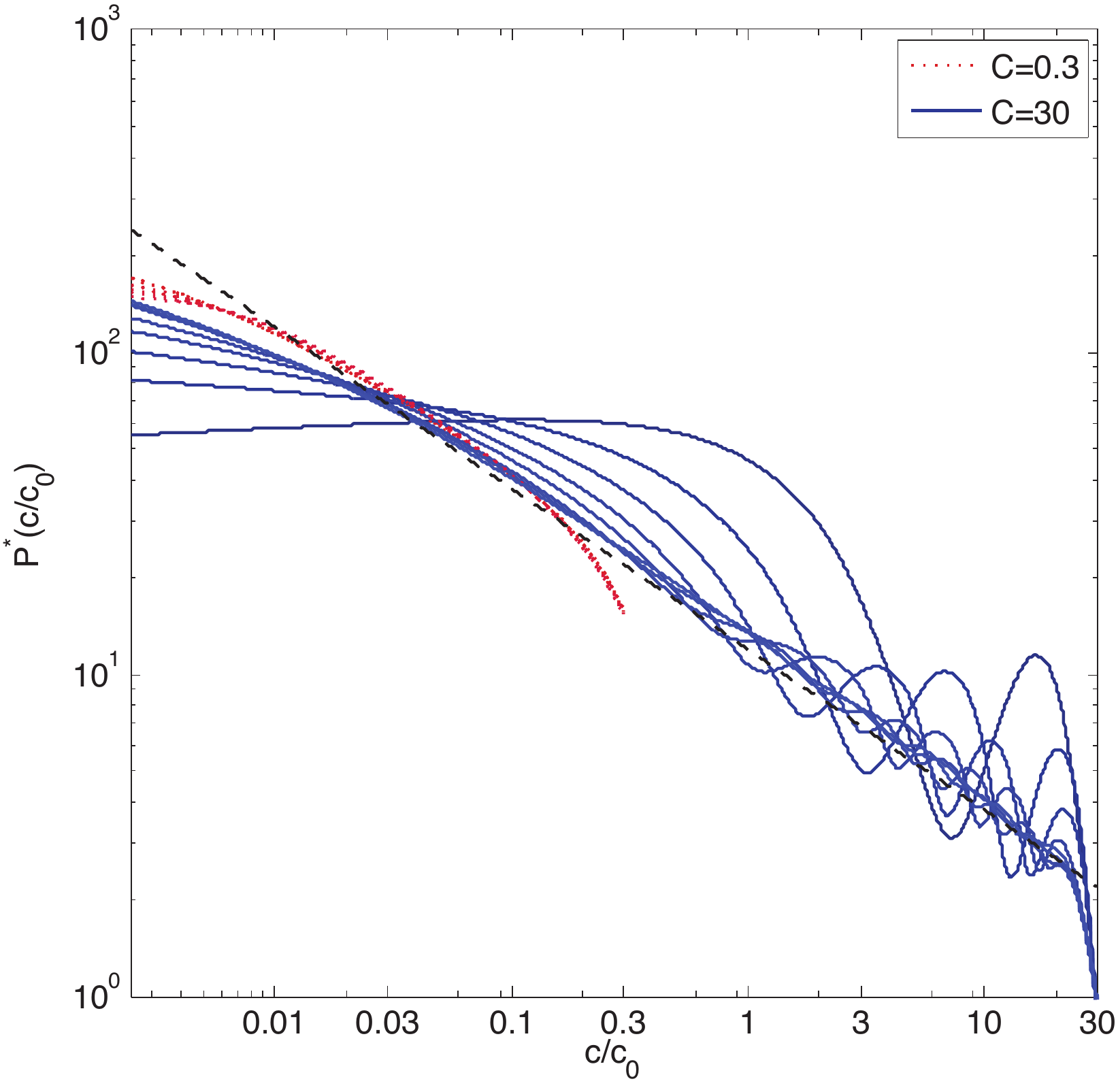}
\caption{(Color online) The optimal probability distribution of inputs, $P_{TF}^*(c)$. In red (dotted line), plotted for $C=0.3$. In blue (solid line), plotted for $C=30$. Different lines correspond to solutions with $2, 3,\cdots, 9$ genes. At low $C$ (red dotted line), the genes are degenerate and the input distribution is independent of the number of genes.  At high $C$ (blue solid line), where the genes tile the concentration range, the optimal input distribution approaches $(c/c_0)^{-1/2}$ (dashed line) as the number of target genes increases. }
\label{noisemanygenes}
\end{figure}

For a system with many target genes, it is illustrative to plot the optimal distribution of input levels, $P^*_{TF}(c)\propto \sigma_c^{-1}(c)$. Figure \ref{noisemanygenes} shows the results for the case of $M=2,3,\cdots,9$ genes at low ($C=0.3$) and high ($C=30$) input dynamic range. At low input dynamic range the distributions for various $M$ collapse onto each other (because the genes are redundant), while at high $C$ increasing the number of genes drives the optimal distribution closer to $\propto c^{-1/2}$.
We recall that the input noise is $\sigma_c \propto \sqrt{c}$, so this shows that, as the number of targets becomes large, the input noise becomes dominant over a wider and wider dynamic range.

\begin{figure*}
\includegraphics[height=3.5in]{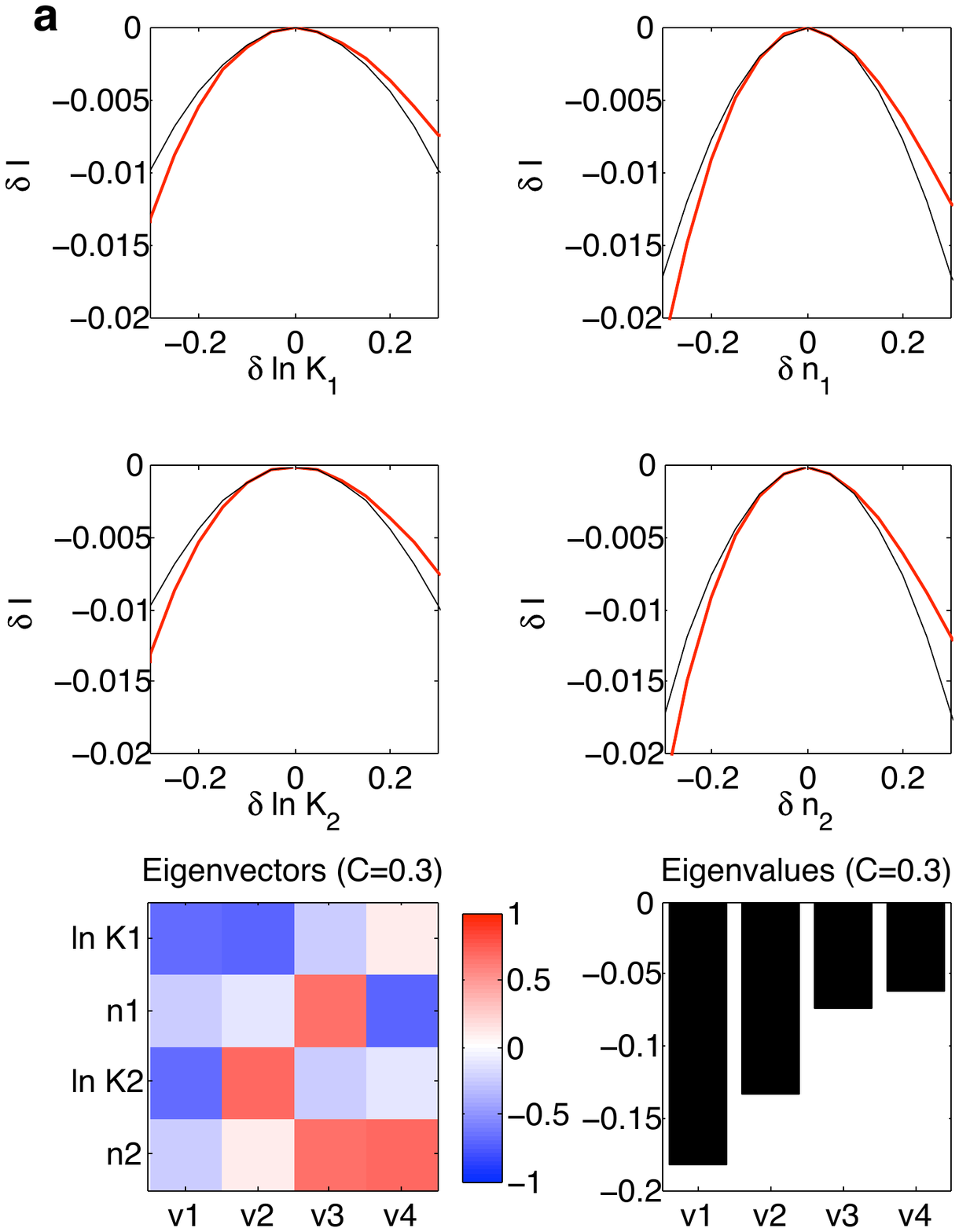}
\includegraphics[height=3.5in]{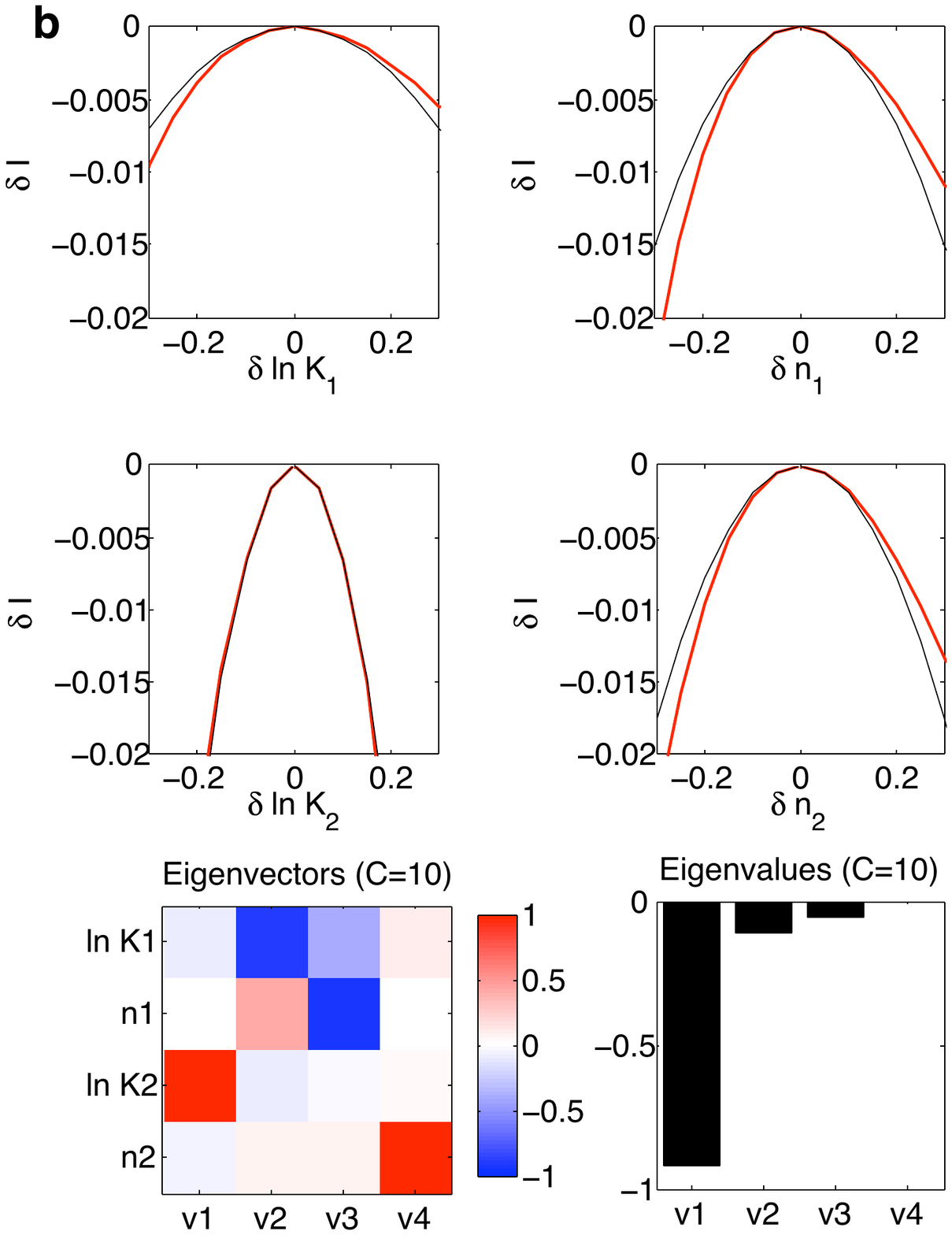}\\
\includegraphics[scale=0.5]{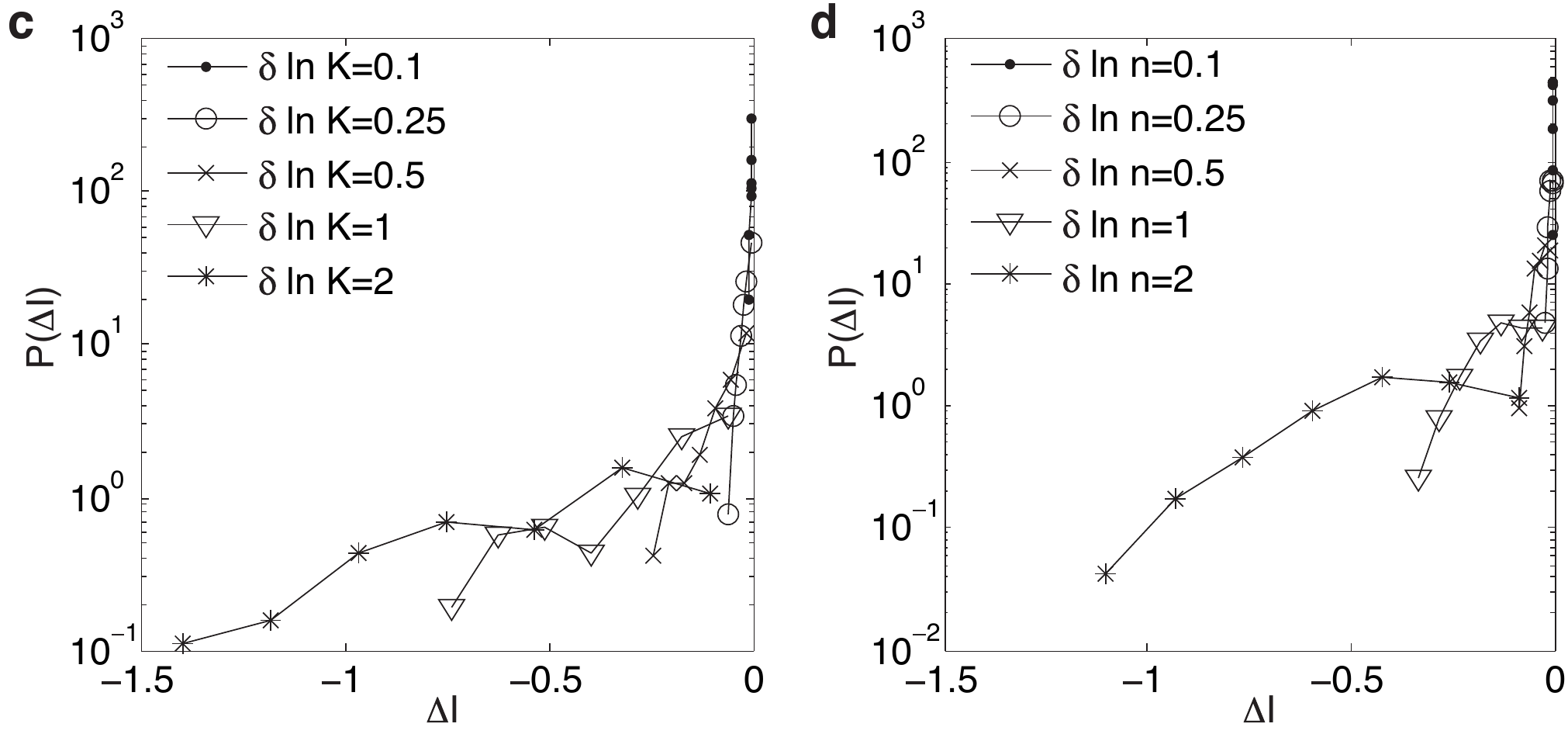}
\caption{(Color online) Parameter variations away from the optimum.  Results are shown for  a two gene system, focusing on the solution with two activators.  (A) Analysis of the Hessian matrix for $c_{\rm max}/c_0 = 0.3$, where the two genes are redundant.  Top four panels show the variation in information ($\delta I$ in bits) along each dimension of the parameter space (thick red line) and the quadratic approximation.  (B) As in (A), but with $c_{\rm max}/c_0 = 10$, where the optimal solution is non--redundnant.
We also show the eigenvectors and eigenvalues of the Hessian matrix.
(C) Distribution of information loss $\Delta I$ when the parameters $K_1$ and $K_2$ are chosen at random from a uniform distribution in $\ln K$, with widths as shown; here  $c_{\rm max}/c_0 = 10$.
(D) As in (C), but for variations in the Hill coefficients $n_1$ and $n_2$.}
\label{Hessian}
\end{figure*}

Finally, one can ask how finely tuned the input/output relations for the particular genes need to be in a maximally informative system. To consider how the capacity of the system changes when the parameters of the input/output relations change slightly, we analyzed the (Hessian) matrix of second derivatives of the  information with respect to fractional changes in the various parameters; we also made more explicit maps of the variations of information with respect to the individual parameters, and sampled the variations in information that result from random variations of the parameters within some range.  Results for  a two gene system  are illustrated in  Fig \ref{Hessian}.

The first point concerns the scale of the variations---$20\%$ changes in parameters away from the optimum result in only $\sim 0.01\,{\rm bits}$ of information loss, and this is true both at low $c_{\rm max}$ where the solutions are redundant and at high $c_{\rm max}$ where they are not.
Interestingly,  the eigenmodes of the Hessian reveal that in the asymmetric case the capacity is most sensitive to variations in the larger $K$. The second most sensitive (much weaker than the first)  direction is a linear combination of both of the parameters $K$ and $n$ for the gene which is activated at lower concentrations.  Perhaps surprisingly, this means that  genes which activate at higher $K$ need to have their input/output relations  positioned with greater accuracy along the $c$ axis, even in fractional terms.  If we think of $K \sim e^{-F/k_B T}$, where $F$ is the binding (free) energy of the transcription factor to its specific target site along the genome, another way of stating this result is that weaker binding energies (smaller $F$) must be specified with greater precision to achieve a criterion level of performance.  
Finally, if we allow parameters to vary at random, we see (Fig \ref{Hessian}C \& D) that the effects on information capacity are extremely small as long as these variations are bounded, so that the range of the natural log of the parameters is significantly less than one.  If we allow larger fluctuations, there is a transition to a much broader distribution of information capacities, with a substantial loss relative to the optimum.

\section{Discussion}

The ability of cells to control the expression levels of their genes is central to growth, development and survival.  In this work we have explored perhaps the simplest model for this control process, in which changes in the concentration of a single transcription factor protein modulate the expression of one or more genes by binding to specific sites along the DNA.  Such models have many parameters, notably the binding energies of the transcription factor to the different target sites and the interactions or cooperativity among factors bound to nearby sites that contribute to the control of the same gene.  This rapid descent from relatively simple physical pictures into highly parameterized models is common to most modern attempts at quantitative analysis of biological systems.  Our goal in this work is to understand whether these many parameters can be determined by appeal to theoretical principles, rather than solely by fitting to data.

We begin our discussion with a caveat.  Evidently,  deriving the many parameters that describe a complex biological system is an ambitious goal, and what we present here is at best a first step.  By confining ourselves to systems in which one transcription factor modulates the expression of many genes, with no further inputs or interactions, we almost certainly exclude the possibility of direct, quantitative comparisons with real genetic control networks.  Understanding this simpler problem, however, is a prerequisite to analysis of more complex systems, and, as we argue here, sufficient to test the plausibility of our theoretical approach.

The theoretical principle to which we appeal is the optimization of information transmission.  In the context of genetic control systems, we can think of information transmission as a measure of control power---if the system can transmit $I\,{\rm bits}$ of information, then adjustment of the inputs allows the cell to access, reliably, $2^I$ distinguishable states of gene expression. In unicellular organisms, for example, these different states could be used to match cellular metabolism to the available nutrients, while in the developing embryo of a multicellular organism these different states could be the triggers for emergence of different cell types or spatial structures; in either case, it is clear that information transmission quantifies our intuition about the control power or (colloquially) complexity that the system can achieve.  Although one could imagine different measures, specialized to different situations, we know from Shannon that the mutual information is the unique measure that satisfies certain plausible conditions, and works in all situations \cite{shannon_48,cover+thomas_91}.

Information transmission is limited by noise.  In the context of genetic control systems, noise is significant because the number of molecules involved in the control process is small, and basic physical principles dictate the random behavior of the individual molecules.  In this sense, the maximization of information transmission really is the principle that organisms should extract maximum control power from a limited number of molecules.  Analysis of experiments on real control elements suggests that the actual number of molecules used by these systems sets a limit of $1-3\,{\rm bits}$ on the capacity of a transcription factor to control the expression level of one gene, that significant increases in this capacity would require enormous increases in the number of molecules, and that, at least in one case, the system can achieve $\sim 90\%$ of its capacity \cite{tkacik+al_08b,tkacik+al_08c}.  Although these observations are limited in scope, they suggest that cells may need to operate close to the informational limits set by the number of molecules that they can afford to devote to these genetic control processes.

The strategy needed to optimize information transmission depends on the structure of the noise in the system.  In the case of transcriptional control, there are two irreducible noise sources, the random arrival of transcription factors at their target sites and the shot noise in the synthesis and degradation of the output molecules (mRNA or protein).  The interplay between these noise sources sets a characteristic scale for the concentration of transcription factors, $c_0 \sim 15 - 150\,{\rm nM}$. If the maximum available concentration is too much larger or smaller than this scale, then the optimization of information transmission becomes degenerate, and we lose predictive power.  Further, $c_0$ sets the scale for diminishing returns, such that increases in concentration far beyond this scale contribute progressively smaller amounts of added information capacity.  Thus, with any reasonable cost for producing the transcription factor proteins,   the optimal tradeoff between bits and cost will set the mean or maximal concentration of transcription factors in the range of $c_0$.  Although only a very rough prediction, it follows without detailed calculation, and it is correct (Table \ref{Cs}).

The optimization of information transmission is largely a competition between the desire to use the full dynamic range of outputs and the preference for  outputs that can be generated reproducibly, that is, at low noise.  Because of the combination of noise sources, this competition has non--trivial consequences, even for a single transcription factor controlling one gene.  As we consider the control of multiple genes, the structure of the solutions becomes richer.  Activators and repressors are both possible, and can achieve nearly identical information capacities.  With multiple target genes, all the combinations of activators and repressors also are possible \cite{mixed_soln}.   This suggests that, generically, there will be exponentially many networks that are local optima, with nearly identical capacities, making it possible for a theory based on optimization to generate diversity.

For a limited range of input transcription factor concentrations, the solutions which optimize information transmission involve multiple redundant target genes.  Absent this result, the observation of redundant targets in real systems  would be interpreted as an obvious sign of non--optimality,  a remnant of evolutionary history, or perhaps insurance against some rare catastrophic  failure of one component. As the available range of transcription factor concentrations becomes larger, optimal solutions diversify, with the responses of the multiple target genes tiling the dynamic range of inputs.  In these tiling solutions, targets that require higher concentrations to be activated or repressed
also are predicted to exhibit greater cooperativity; in such an optimized system one thus should find some genes controlled by a small number of strong binding sites for the transcription factor, and other genes with a large number of weaker sites.

To a large extent, the basic structure of the (numerically) optimal solutions can be recovered analytically through various approximation schemes.  These analytic approximations make clear that the optimization really is driven by a conflict between using the full dynamic range of outputs and avoiding states with high intrinsic noise.  In particular, this means that simple intuitions based on maximizing the entropy of output states, which are correct when the noise is unstructured \cite{laughlin_81}, fail.  Thus, almost all solutions have the property that at least one target gene is not driven through the full dynamic range of its outputs, and even with one gene the midpoint of the optimal activation curve can be far from the midpoint of the available range of inputs.  The interplay between different noise sources also breaks the symmetry between activators and repressors, so that repressors optimize their information transmission by using only a small fraction of the available input range.

The predictive power of our approach depends on the existence of well defined optima.  At the same time, it would be difficult to imagine evolution tuning the parameters of these models with extreme precision, so the optima should not be too sharply defined.  Indeed, we find that optima are clear but broad.  In the case of multiple genes,  random $\sim 25\%$ variations in parameters around their optima result in only tiny fractions of a bit of information loss, but once fluctuations become larger than this the information drops precipitously.  Looking more closely, we find that proper placement of the activation curves at the upper end of the input range is more critical, implying that it is actually the weaker binding sites whose energies need to be adjusted more carefully (perhaps contrary to intuition).

With modest numbers of genes, the optimization approach we propose here has the promise of making rather detailed predictions about structure of the input/output relations, generating what we might think of as a spectrum of $K$s and $n$s. In the limit of larger networks, we might expect this spectrum to have some universal properties, and we see hints of this in Fig \ref{noisemanygenes}.  Here, as we add more and more target genes, the optimal distribution of inputs approaches an asymptote $P_{TF}(c) \propto 1/\sqrt{c}$; more of this limiting behavior is accessible if the available dynamic range of inputs is larger.  This is the form we expect if the effective noise is dominated by the input noise, $\sigma_c \propto \sqrt{c}$.  Thus, adding more targets and placing them optimally allows the system to suppress output noise and approach ever more closely the fundamental limits set by the physics of diffusion.

Although there are not so many direct physical measurements specifying the input/output relations of genetic regulatory elements, there are many systems in which there is evidence for  `tiling' of the concentration axis by a set of target genes, all regulated by the same transcription factor, along the lines predicted here  \cite{timing}. For example, in quorum sensing by bacteria, the concentrations of extracellular signaling molecules are translated internally into different concentrations of LuxR, which acts as a transcription factor on a number of genes, and these can be classified as being responsive  to low, intermediate and high levels of LuxR 
\cite{waters+al_06}.  Similarly, the decision of {\it Bacillus subtilis} to sporulate is controlled by the phosphorylated form of the transcription factor Spo0A, which regulates the expression of $\sim 30$ genes as well as an additional 24 multi--gene operons \cite{molle+al_03}.  For many of these targets the effects of SpoA$\sim$P are direct, and the sensitivity to high vs low concentrations can be correlated with the binding energies of the transcription factor to the particular promoters \cite{fujita+al05}.  In yeast, the transcription factor Pho4 is a key regulator of phosphate metabolism, and activates targets such as {\em pho5} and {\em pho84} at different concentrations \cite{lam+al_06}.  All of these are potential test cases for the theoretical approach we have outlined here (each with its own complications), but a substantially new level of quantitative experimental work would be required to test the theory meaningfully.

The classic example of multiple thresholds in the activation of genes by a single transcription factor is in embryonic development \cite{wolpert_69,lawrence_92}. In this context, spatial gradients in the concentration of transcription factors and other signaling molecules mean that otherwise identical cells in the same embryo experience different inputs.  If multiple genes are activated by the same transcription factor but at different thresholds, then smooth spatial gradients can be transformed into sharper `expression domains' that provide the scaffolding for more complex spatial patterns.  Although controversies remain about the detailed structure of the regulatory network, the control of the `gap genes' in the {\em Drosophila} embryo by the transcription factor Bicoid seems to provide a clear example of these ideas \cite{lawrence_92,driever+nv_88a,driever+nv_88b,driever+nv_89,struhl+al_89,burz+al_98}.  Recent experimental work \cite{gregor+al_07a,gregor+al_07b} suggests that it will be possible to make absolute measurements of (at least) Bicoid concentrations, and to map the input/output relations and noise in this system, holding out the hope for more quantitative comparison with theory.

Finally, we look ahead to the more general problem in which multiple target genes are allowed to interact.  Absent these interactions, even our optimal solutions have a strong degree of redundancy---as the different targets turn on at successively higher concentrations of the input, there is a positive correlation and hence redundancy among the signals that they convey.  This redundancy could be removed by mutually repressive interactions among the target genes,   increasing the efficiency of information transmission in much the same way that lateral inhibition or center--surround organization enhances the efficiency of neural coding in the visual system \cite{barlow_59,atick+redlich_90}.  It is known that such mutually repressive interactions exist, for example among the gap genes in the {\em Drosophila} embryo \cite{repress}.  The theoretical challenge is to see if these observed structures can be derived, quantitatively, from the optimization of information transmission.  

\begin{acknowledgments}
We thank T Gregor, JB Kinney, P Mehta, T Mora, SF N{\o}rrelykke, ED Siggia, and especially CG Callan for helpful discussions.  Work at Princeton was supported in part by NSF Grant PHY--0650617, and by NIH Grants P50 GM071508 and R01 GM077599.  GT was supported in part by NSF grants DMR04--25780, IBN-0344678, and by the Vice Provost for Research at the University of Pennsylvania.    WB also thanks his colleagues at the University of Rome, La Sapienza, for their hospitality during a portion of this work.
\end{acknowledgments}


\begin{thebibliography}{99}
\bibitem{remark}
Strictly speaking, Shannon considered the question ``how much does the output tell us about the input?'', which is almost the same as the question about control power posed here.
\bibitem{note_Gauss}
In general, if we allow for interactions,
\begin{widetext} 
\begin{equation}
P(\{g_{\rm i}\}| c)  = {1\over {(2\pi )^{M/2}}} \exp\left[ 
{1\over 2} \ln \det( {\sigma}) - {1\over 2} \sum_{{\rm i, j}=1}^M
\left( g_{\rm i} - \bar g_{\rm i}(c)\right) {\cal K}_{\rm ij} 
\left( g_{\rm j} - \bar g_{\rm j}(c)\right)
\right],
\end{equation}
\end{widetext}
 where $\cal K$ measures the (inverse) covariance of the fluctuations or noise in the expression levels at fixed input, 
 \begin{equation}
 \langle \left( g_{\rm i} - \bar g_{\rm i}(c)\right) \left( g_{\rm j} - \bar g_{\rm j}(c)\right)\rangle = ({\cal K}^{-1})_{\rm ij} .
 \end{equation}
In the present paper we restrict the discussion to independent genes ${\cal K}_{\rm ij} = {{\delta_{\rm ij}}/{\sigma_{\rm i}^2}}$.
\bibitem{continuous}
We are being a bit sloppy here, since the input $c$ is a continuous variable, but the intuition is correct.
\bibitem{note1}
This normalization of the mean expression levels isn't exactly the same as fixing an absolute maximum, but in the small noise limit the difference is not significant.
\bibitem{remark2}
$N_{\rm max}$ is the maximum number of independent molecular events, which is not the same as the maximum number of protein molecules expressed, but they are proportional.
\bibitem{mixed_soln}
Most transcription factors act either as activators or repressors, but not both.  Thus, we should not take the existence of locally optimal solutions with `mixed' functions for the input molecule too literally.  We think the important lesson here is that even simple formulations of our optimization problem have many solutions, so we expect that a realistic formulation, with multiple interacting targets, will also have multiple solutions.  This is crucial if we wnata theory based on optimization to predict the diversity of biological mechanisms.
\bibitem{timing}
Our discussion here has been in the limit where expression levels are in steady state.  A complementary view of why tiling of the concentration axis is important is that, as the transcription factor concentration changes over time, the cell would cross the thresholds for activation or repression of different genes in sequence, generating a temporal program of expression in which different proteins are synthesized at different times.  See, for example, Refs \cite{ronen+al_02} and \cite{zaslaver+al_04}.
\bibitem{repress}
For a sampling of the evidence and models that describe these interactions, see Refs \cite{lawrence_92,jackle+al_86,rivera-pomar+jackle_96,sanchez+thieffry_01,yu+small_08}.
\end{thebibliography}
\end{document}